%% file: iclr2024_conference.tex
\documentclass{article} %
\usepackage{iclr2024_conference,times}

\input{macros}

\usepackage{hyperref}
\usepackage{url}

\title{Who Leaked the Model? Tracking IP Infringers in Accountable Federated Learning}

\author{Shuyang Yu\textsuperscript{1}, Junyuan Hong\textsuperscript{1,2}, 
Yi Zeng\textsuperscript{3}, 
Fei Wang\textsuperscript{4}, Ruoxi Jia\textsuperscript{3} and Jiayu Zhou\textsuperscript{1} \\
       \textsuperscript{1}Department of Computer Science and Engineering, Michigan State University\\
       \textsuperscript{2}Department of Electrical and Computer Engineering, University of Texas at Austin \\
       \textsuperscript{3}Department of Computer Engineering, Virginia Tech \\
       \textsuperscript{4}Department of Population Health Sciences, Cornell University \\
       \texttt{\{yushuyan,hongju12,  jiayuz\}@msu.edu}, \texttt{\{yizeng,ruoxijia\}@vt.edu},\\ \texttt{few2001@med.cornell.edu}
       }

\iclrfinalcopy %
\begin{document}

\maketitle

\begin{abstract}
Federated learning (FL) emerges as an effective collaborative learning framework to coordinate data and computation resources from massive and distributed clients in training.
Such collaboration results in non-trivial intellectual property (IP) represented by the model parameters that should be protected and shared by the whole party rather than an individual user.
Meanwhile, the distributed nature of FL endorses a malicious client the convenience to compromise IP through illegal model leakage to unauthorized third parties.
To block such IP leakage, it is essential to make the IP identifiable in the shared model and locate the anonymous infringer who first leaks it.
The collective challenges call for \emph{accountable federated learning}, which requires verifiable ownership of the model and is capable of revealing the infringer's identity upon leakage. 
In this paper, we propose
Decodable Unique Watermarking (DUW) for complying with the requirements of accountable FL.
Specifically, before a global model is sent to a client in an FL round, DUW encodes a client-unique key into the model by leveraging a backdoor-based watermark injection.
To identify the infringer of a leaked model, DUW examines the model and checks if the triggers can be decoded as the corresponding keys.
Extensive empirical results show that DUW is highly effective and robust, achieving over $99\%$ watermark success rate for Digits, CIFAR-10, and CIFAR-100 datasets under heterogeneous FL settings, and identifying the IP infringer with $100\%$ accuracy even after common watermark removal attempts.
\end{abstract}
\input{sec/intro}
\input{sec/related}

\input{sec/method}

\input{sec/exp}

\vspace{-0.1in}
\section{Conclusion}
In this paper, 
we target at accountable FL, and propose
Decodable Unique Watermarking (DUW), that can verify the FL model's ownership and track the IP infringers in the FL system at the same time. Specifically, the server will embed a client-unique key into each client's local model before broadcasting. The IP infringer can be tracked according to the decoded keys from the suspect model. Extensive experimental results show the effectiveness of our method in accurate IP tracking, confident verification, model utility preserving, and robustness against various watermark removal attacks.

\section*{Acknowledgement}
This research was supported by the National Science Foundation (IIS-2212174, IIS-1749940), National Institute of Aging (IRF1AG072449).

\bibliography{auto_gen}
\bibliographystyle{iclr2024_conference}

\clearpage

\appendix

\input{sec/appendix}

\end{document}

%% file: macros.tex
\usepackage{tablefootnote}
\usepackage{amsthm}
\usepackage{amssymb}
\usepackage{mathtools}
\usepackage{hyperref}
\usepackage[nameinlink,capitalize]{cleveref}
\hypersetup{colorlinks=true,linkcolor=blue,citecolor=blue,urlcolor=blue,pdfborder={0 0 0}}
\usepackage[normalem]{ulem} %
\usepackage{mathtools}
\usepackage{algorithm,algorithmic}

\usepackage[utf8]{inputenc} %
\usepackage[T1]{fontenc}    %
\usepackage{hyperref}       %
\usepackage{url}            %
\usepackage{booktabs}       %
\usepackage{amsfonts}       %
\usepackage{nicefrac}       %
\usepackage{microtype}      %
\usepackage{color, colortbl}
\definecolor{Gray}{gray}{0.9}
\newcolumntype{g}{>{\columncolor{Gray}}c}

\usepackage{graphicx,wrapfig}
\usepackage{caption}
\usepackage{subcaption}
\usepackage{amsfonts}
\usepackage{url}
\usepackage{enumitem}
\usepackage{amsthm}
\usepackage{thmtools}
\usepackage{thm-restate}

\theoremstyle{definition}
\newtheorem{definition}{Definition}[section]
\theoremstyle{remark}

\newcommand{\vtheta}{{\boldsymbol{\theta}}}

\newcommand{\cA}{{\mathcal{A}}}

\newcommand{\cD}{{\mathcal{D}}}

\newcommand{\cX}{{\mathcal{X}}}
\newcommand{\cY}{{\mathcal{Y}}}
\newcommand{\cZ}{{\mathcal{Z}}}

\DeclareMathOperator*{\argmax}{arg\,max}

\newcommand{\bc}{\begin{center}}
\newcommand{\ec}{\end{center}}

\newcommand{\bdm}{\begin{displaymath}}
\newcommand{\edm}{\end{displaymath}}

\newcommand{\beq}{\begin{equation}}
\newcommand{\eeq}{\end{equation}}

\newcommand{\bfl}{\begin{flushleft}}
\newcommand{\efl}{\end{flushleft}}

\newcommand{\bt}{\begin{tabbing}}
\newcommand{\et}{\end{tabbing}}

\newcommand{\beqn}{\begin{align}}
\newcommand{\eeqn}{\end{align}}

\newcommand{\beqs}{\begin{align*}} %
\newcommand{\eeqs}{\end{align*}}  %

\renewcommand{\algorithmicrequire}{\textbf{Input:}}

%% file: sec/intro.tex
\section{Introduction}
\label{sec:intro}
Federated learning (FL)~\citep{konevcny2015federated} has been widely explored as a distributed learning paradigm to enable remote clients to collaboratively learn a central model without sharing their raw data, effectively leveraging the massive and diverse data available in clients for learning and protecting the data confidentiality. 
The learning process of FL models typically requires the coordination of significant computing resources from a multitude of clients to curate the valuable information in the client's data, and the FL models usually have improved performance than isolated learning and thus high commercial value. 
Recently, the risk of leaking such high-value models has drawn the attention of the public. One notable example is the leakage of the foundation model from Meta~\citep{meta2023} by users who gained the restricted distribution of models.
The leakage through restricted distribution could be even more severe in FL which allows all participating clients to gain access to the valued model.
For each iterative communication round, a central server consolidates models from various client devices, forming a global or central model. This model is then disseminated back to the clients for the next update, and therefore the malicious clients have full access to the global models. 
As such, effectively protecting the global models in FL is a grand challenge. 

\begin{figure}
    \centering
    \includegraphics[width=13.8cm]{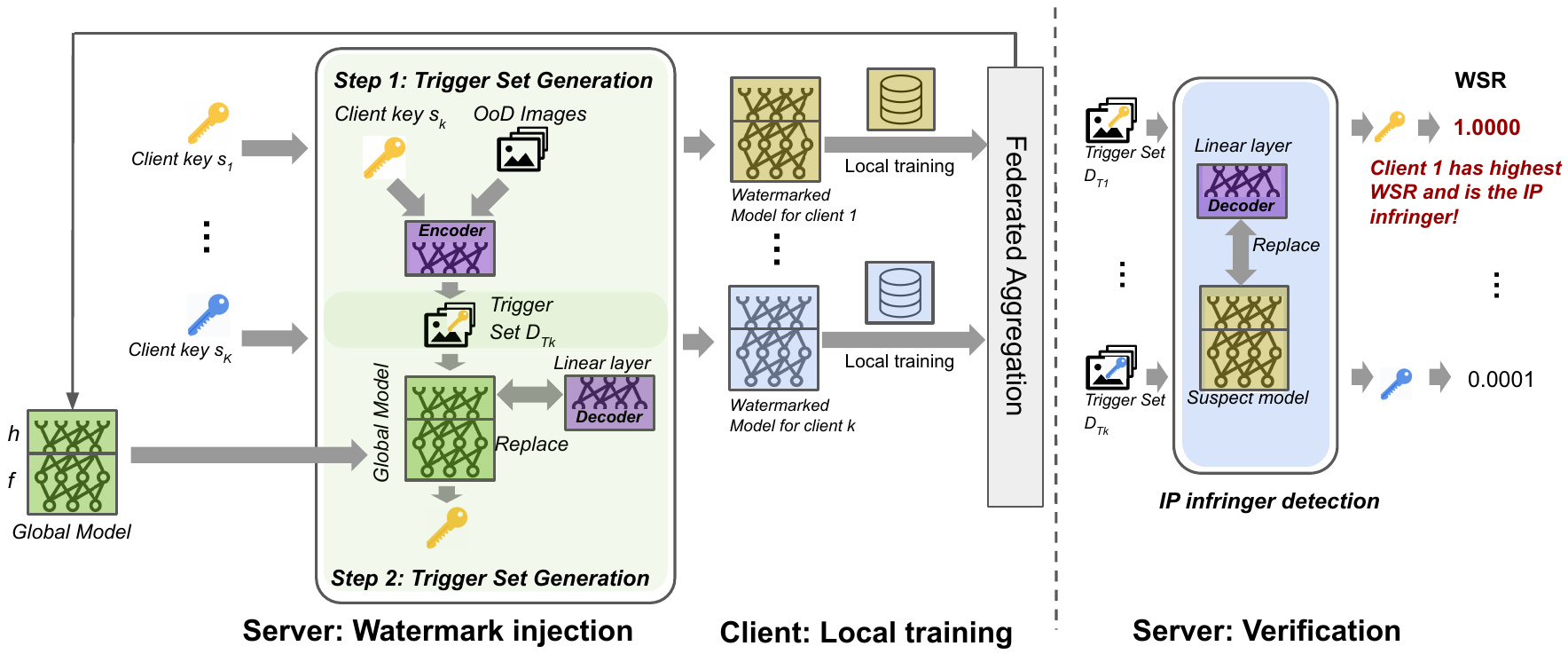}
    \vspace{-0.1in}
    \caption{
    The proposed Decodable Unique Watermarking (DUW) for watermark injection and verification. During watermark injection, the server first uses client-unique keys and an OoD dataset as the input for the pre-trained encoder to generate trigger sets. When the server implants the watermark based on the objective function $J'(\theta_k)$~(\cref{eq:l2}), a decoder is utilized to replace the classifier head. During verification, the suspect model is tested on all the trigger sets, and the client that leaked the model is identified as the one that achieves the highest WSR~(\cref{eq:wsr}) in trigger sets.}
    \label{fig:framework}
    \vspace{-0.3in}
\end{figure}

Watermarking techniques~\citep{adi2018turning,chen2021you,darvish2019deepsigns,fan2019rethinking,uchida2017embedding,zhang2018protecting} are recently introduced to verify the IP ownership of models. 
Among them, backdoor-based watermarking shows strong applicability because of its model-agnostic nature, which repurposes the backdoor attacks of deep models and uses special-purposed data (trigger set) to insert hidden patterns in the model to produce undesired outputs given inputs with triggers~\citep{zhang2018protecting,le2020adversarial,goldblum2022dataset,li2022leveraging}. 
A typical backdoor-based watermarking operates as follows:
The model owner first generates a trigger set consisting of samples paired with pre-defined target labels. 
The owner then embeds the watermark into the model by fine-tuning the model with the trigger set and the original training samples. 
To establish the ownership of the model, one evaluates the accuracy of the suspect model using the trigger set. 
The mechanism safeguards the assumption that only the watermarked model would perform exceptionally well on the unique trigger set. 
If the model's accuracy on the trigger set surpasses a significant threshold, the model likely belongs to the owner. 

Conventional backdoor-based watermarking, however, does not apply to FL settings because of the required access to the training data to maintain model utility.
To address the challenge, \cite{tekgul2021waffle} proposed WAFFLE, which utilized only random noise and class-consistent patterns to embed a backdoor-based watermark into the FL model. However, since WAFFLE injected a unified watermark for all the clients, it cannot solve another critical question:\emph{
Who is the IP infringer among the FL clients?}
Based on WAFFLE, \cite{shao2022fedtracker} introduced a two-step method
 FedTracker
to verify the ownership of the model with the central watermark from WAFFLE, and
track the malicious clients in FL by embedding unique local fingerprints into local models. However, the local fingerprint in~\cite{shao2022fedtracker} is a parameter-based method, which is not applicable for many practical scenarios, where many re-sale models are reluctant to expose their parameters, and the two-step verification is redundant. 
Therefore, how to spend the least effort on changing the model while verifying and 
tracking the IP infringers using the same watermark in FL remains to be a challenging problem.

The aforementioned challenges call for a holistic solution towards \emph{accountable federated learning}, which is characterized by the following essential requirements: \underline{R1)~Accurate IP tracking:} Each client has a unique ID to trace back. IP tracking should be confident to identify one and only one client. 
\underline{R2)~Confident verification:} The ownership verification should be confident. \underline{R3)~Model utility:} The watermark injected should have minimal impact on standard FL accuracy. 
\underline{R4)~Robustness:} The watermark should be robust and resilient against various watermark removal attacks. 
In this paper, we propose a practical watermarking framework for FL called Decodable Unique Watermarking (DUW) to comply with these requirements. 
Specifically, we first generate unique trigger sets for each client by using a pre-trained encoder~\citep{li2021invisible} to embed client-wise unique keys to one randomly chosen out-of-distribution (OoD) dataset. 
During each communication round, the server watermarks the aggregated global model using the client-wise trigger sets before dispatching the model. 
A decoder replaces the classifier head in the FL model during injection so that we can decode the model output to the client-wise keys. 
We propose a regularized watermark injection optimization process to preserve the model's utility.
During verification, the suspect model is tested on the trigger sets of all the clients, and the client that achieves the highest watermark success rate (WSR) is considered to be the IP infringer. The framework of method is shown in \cref{fig:framework}.

The contributions of our work can be summarized in three folds: \\
    $\bullet$ We make the FL model leakage from anonymity to accountability by injecting
    DUW. DUW enables ownership verification and leakage tracing at the same time without access to model parameters during verification. \\
    $\bullet$ With utility preserved, both the ownership verification and IP tracking of our DUW are not only accurate but also confident without collisions. \\
    $\bullet$ Our DUW is robust against existing watermarking removal attacks, including fine-tuning, pruning, model extraction, and parameter perturbation.

%% file: sec/related.tex
\section{Related Work and Background}
\label{sec:related}

\textbf{Federated learning (FL)} is a distributed learning framework that enables massive and remote clients to collaboratively train a high-quality central model~\citep{konevcny2016federated}. 
This paper targets the cross-silo FL with at most hundreds of clients~\citep{marfoq2020throughput}. In the cross-silo setting, each client is an institute, like a hospital or a bank. 
It is widely adopted in practical scenario~\citep{ bagdasaryan2020backdoor,t2020personalized,zhu2021data,tekgul2021waffle}. %
FedAvg~\citep{mcmahan2017communication} is one of the representative methods for FL, which averages local models during aggregation. This work is based on the FedAvg. Suppose we have $K$ clients, 
and our FL model $M$ used for standard training consists of two components, including a feature extractor $f:\cX \to \cZ$ governed by ${\theta^f}$, and a classifier $h:\cZ \to \cY$ governed by ${\theta^h}$, where $\cZ$ is the latent feature space. 
The collective model parameter is $\theta=(\theta^h,\theta^f)$. The objective for a client's local training is: 
\begin{align}\label{eq:local_train}
    J_{k}(\theta):=\frac{1}{|\cD_k|}\sum\nolimits_{(x,y)\in \cD_{k}}\ell(h(f(x;\theta^f);\theta^h), y),
\end{align}
where $\cD_k$ is the local dataset for client $k$, 
and $\ell$ is the cross-entropy loss.
The overall objective function of FL is thus given by
 $\min_{\vtheta} \frac{1}{K} \sum \nolimits_{k=1}^K J_{k}(\theta)$.

\textbf{DNN watermarking}
can be categorized into two main streams: parameter-based watermarking and backdoor-based watermarking.
\\\emph{Parameter-based watermarking} approaches~\citep{darvish2019deepsigns,uchida2017embedding,kuribayashi2021white,mehta2022aime} embed a bit string as the watermark into the parameter space of the model. 
The ownership of the model can be verified by comparing the watermark extracted from the parameter space of the suspect model and the owner model. \cite{shao2022fedtracker} proposed a parameter-based watermarking method for FL called FedTracker. 
It inserts a unique parameter-based watermark into the models of each client to verify the ownership.
However, all parameter-based watermarking requires an inspection of the parameters of the suspect models, which is not applicable enough for many re-sale models.
\\\emph{Backdoor-based watermarking}~\citep{zhang2018protecting,le2020adversarial,goldblum2022dataset,li2022leveraging} does not require access to model parameters during verification. 
The watermark is embedded by fine-tuning the model with a trigger set $\cD_T$ and clean dataset $\cD$. 
Pre-defined target label $t$ is assigned to $\cD_T$. 
The objective for the backdoor-based watermarking is formulated as:
\begin{align}\label{eq:normal_inject}
    J(\theta):=\frac{1}{|\cD|}\sum\nolimits_{(x,y)\in \cD} \ell(h(f(x;\theta^f);\theta^h),y)+\frac{1}{|\cD_{T}|}\sum \nolimits_{(x,t)\in \cD_{T}}\ell(h(f(x;\theta^f);\theta^h), t),
\end{align}
Upon verification, we verify the suspect model $M_s$ on the trigger set $\cD_T$. 
If the accuracy of the trigger set is larger than a certain threshold $\sigma$, the ownership of the model can be established. 
We formally define the ownership verification of the backdoor-based model as follows:
\begin{definition}[Ownership verification]\label{def:verification}
    We define watermark success rate (WSR) as the accuracy on the trigger set $\cD_{T}$: 
\begin{equation}\label{eq:wsr}
        \text{WSR} = Acc(M_s, \cD_{T}).
    \end{equation}
If $\text{WSR}>\sigma$, the ownership of the model is established.
\end{definition}
WAFFLE~\citep{tekgul2021waffle} is the first FL backdoor-based watermarking, which utilized random noise and class-consistent patterns to embed a backdoor-based watermark into the FL model. However, WAFFLE can only verify the ownership of the model, yet it
cannot track the specific IP infringers. %

%% file: sec/method.tex
\section{Method}
\label{sec:method}

Watermarking has shown to be a feasible solution for IP verification, and the major goal of this work is to seek a powerful extension for traceable IP verification for accountable FL that can accurately identify the infringers among a scalable number of clients.
A straightforward solution is injecting different watermarks for different clients.
However, increasing the number of watermarks could lower the model's utility as measured by the standard accuracy due to increased forged knowledge %
~\citep{tang2020embarrassingly} (\underline{R3}).
Meanwhile, maintaining multiple watermarks could be less robust to watermark removal because of the inconsistency between injections (\underline{R4}).
Accurate IP tracking (\underline{R1}) is one unique requirement we seek to identify the infringer's identity as compared with traditional watermarking in central training. 
The greatest challenge in satisfying \underline{R1} is addressing the watermark \emph{collisions} between different clients. 
A watermark collision is when the suspect model produces similar watermark responses on different individual verification datasets in FL systems. Formally:
\begin{definition}[Watermark collision]\label{def:wm_coll}
      During verification in \cref{def:verification}, we test the suspect model $M_s$ on all the verification datasets $\cD_T=\{ \cD_{T_1}, \dots, \cD_{T_k}, \dots, \cD_{T_{K}}\}$ of all the clients to identify the malicious client, and WSR for the $k$-th  verification datasets is defined as $\text{WSR}_k$.
      If we have 
multiple clients $k$ satisfying
      $\text{WSR}_k=Acc(M_s, \cD_{T_k})>\sigma$, the ownership of suspect model $M_s$ can be claimed for more than one client, then the watermark collisions happen between clients.
\end{definition}
\subsection{Pitfalls for Watermark Collision}\label{sec:pitfall}
To avoid watermark collision, one straightforward solution is to simply design different trigger sets for different clients. However, this strategy may easily lead to the watermark-collision pitfall. 
We use traditional backdoor-based watermarking by adding 
 arbitrary badnet~\citep{gu2019badnets} triggers using random noise or 0-1 coding trigger for each client as examples to demonstrate this pitfall.
We conduct the experiments on CIFAR-10 with 100 clients, during $4$ injection rounds, at least $89\%$ and $87\%$ of the clients have watermark collisions for two kinds of triggers, respectively. 

To analyze why these backdoor-based watermarkings lead us into the trap,
we list all the clients with watermark collisions for one trial, and define the client\_{ID} with the highest WSR as the predicted client\_{ID}. We found that $87.5\%$ of the predicted client\_{ID} share the same target label as the ground truth client, and for the rest $12.5\%$ clients, both the trigger pattern and target label are different. %
Based on the results, 
we summarize two possible reasons:
1) The same target labels will easily lead to the watermark collision.
2) The trigger pattern differences between clients are quite subtle, so the differences between the watermarked models for different clients are hard to detect.  
Thus, in order to avoid this pitfall, we have to ensure the uniqueness of both the triggers and target labels between different clients.  More experiment settings and results for pitfalls can be referred to \cref{sec:baseline}.

\subsection{Decodable Unique Watermarking}\label{sec:uwatermark}
In this section, we propose the Decodable Unique Watermark (DUW) that can simultaneously address the four requirements of accountable FL summarized in \cref{sec:intro}: \underline{R1 (accurate IP tracking)}, \underline{R2 (confident verification)}, \underline{R3 (model utility)}, \underline{R4 (robustness)}. 
In DUW, all the watermarking is conducted on the server side, so no computational overhead is introduced to clients.
Before broadcasting the global model to each local client, the server will inject a unique watermark for each client. The watermark is unknown to clients but known to the server (see \cref{fig:framework} server watermark injection). 
Our DUW consists of the following two steps for encoding and decoding the client-unique keys.

\textbf{Step 1: Client-unique trigger encoding.}
Due to the data confidentiality of FL, the server 
has no access to any data from any of the clients. Therefore for watermark injection, the server needs to collect or synthesize some OoD data for trigger set generation. The performance of the watermark is not sensitive to the choice of the OoD datasets. 

To accurately track the malicious client, we have to distinguish between watermarks for different clients.
High similarity between trigger sets of different clients is likely to cause watermark collisions among the clients (see \cref{sec:pitfall}), which makes it difficult to identify which client leaked the model. 

To solve this problem,
 we propose to use a pre-trained encoder $E: \cX \rightarrow \cX$ governed by $\theta_E$ from~\cite{li2021invisible} to generate unique trigger sets for each client. 
This backdoor-based method provides a successful injection of watermarks with close to $100\%$ WSR, which ensures the confident verification (\underline{R2}).
We design a unique key corresponding to each client ID as a one-hot binary string to differentiate clients. 
For instance, for the $k$-th client, the $k$-th entry of the key string $s_k$ is $1$, and the other entries are $0$. 
We set the length of the key as $d$, where  $d \geq K$.
For each client, the key can then be embedded into the sample-wise triggers of the OoD samples by feeding the unique key and OoD data to the
pre-trained encoder. 
The output of the encoder makes up the trigger sets. 
The trigger set for the $k$-th client is defined in 
$
    \cD_{T_k}=\{(x',t_k)|x' \sim E_{x\in \cD_{OoD}}(x, s_k;\theta_E)\}$,
where $\cD_{OoD}$ is a randomly chosen OoD dataset, and $t_k$ is the target label for client $k$.
To this end, different trigger sets for different clients will differ by their unique keys, and watermark collision can be alleviated (\underline{R1}). Note that our trigger sets will be the same as verification datasets.

\textbf{Step 2: Client-unique target label by decoding triggers to client keys.}
The main intuition is that the same target label of the trigger sets may still lead to watermark collisions even if the keys are different (see \cref{sec:pitfall}).
Thus, we propose to project the output dimension of the original model $M$ to a higher dimension, larger than the client number $K$, to allow each client to have a unique target label. 
To achieve this goal, we first set the target label $t_k$ in the trigger set $\cD_{T_k}$ to be the same as the input key $s_k$ corresponding to each client, and then use a decoder $D: \cZ \rightarrow \cY$ parameterized by $\theta_D$ to replace the classifier $h$ in the FL training model $M$. 
The decoder $D$ only has one linear layer, whose input dimension is the same as the input dimension of $h$, and its output dimension is the length of the key. 
To avoid watermark collision between clients induced by the target label, we make the decoder weights orthogonal with each other during the random initialization so that the watermark injection tasks for each client can be independent (\underline{R1}). 
The weights of the decoder are frozen once initialized to preserve the independence of different watermark injection tasks for different clients.
Suppose ${\theta_k=(\theta_k^f, \theta_k^h)}$ is the parameter which will be broadcast for client $k$, 
we formulate the injection optimization as:
\begin{align}\label{eq:inject}
\min_{\theta^f_k}J(\theta^f_k):=\frac{1}{|\cD_{T_k}|} \sum \nolimits_{(x',s_k)\in \cD_{T_k}}\ell(D(f(x';\theta^f_k);\theta_D), s_k),
\end{align}
The classifier $h$ will be plugged back into the model before the server broadcasts the watermarked models to clients.
Compared with traditional backdoor-based watermarking (\cref{eq:normal_inject}), our watermark injection requires no client training samples, 
which ensures the data confidentiality of FL.

\textbf{Robustness.} Our framework also brings in robustness against fine-tuning-based watermark removal (\underline{R4}). The main intuition is that replacing classifier $h$ with decoder $D$ also differs the watermark injection task space from the original classification task space. 
Since the malicious clients have no access to the decoder and can only conduct attacks on model $M$, the attacks have more impact on the classification task instead of our watermark injection task, which makes our decodable watermark more resilient against watermark removal attacks.

\subsection{Injection Optimization with Preserved Utility}\label{sec:norm}
While increasing the size of the client number, watermark injection in the OoD region may lead to a significant drop in the standard FL accuracy (\underline{R3}) because of the overload of irrelevant knowledge. 
An ideal solution is to bundle the injection with training in-distribution (ID) data, which however is impractical for a data-free server.
Meanwhile, lacking ID data to maintain the standard task accuracy, the distinct information between the increasing watermark sets and the task sets could cause the fade-out of the task knowledge.
We attribute such knowledge vanishing to the divergence in the parameter space between the watermarked and the original models.
Thus, we propose to augment the injection objective \cref{eq:inject} with a $l_2$ regularization on the parameters:
\begin{align}\label{eq:l2}
    \min_{\theta_k} J'(\theta^f_k):=J(\theta^f_k)+ \frac{\beta}{2} \|\theta^f_k - \theta^f_g\|^2,
\end{align}
where $\theta^f_g$ is the original parameter of the global model. 
The regularization term of \cref{eq:l2} is used to restrict the distance between the watermarked model and the non-watermarked one so that the utility of the model can be better preserved (\underline{R3}). 
Our proposed DUW 
is summarized in \cref{algorithm:inject}.

\subsection{Verification}
During verification, we not only verify whether the suspect model $M_s=(f_s,h_s)$ is a copy of our model $M$, but also track who is the leaker among all the clients by examining if the triggers can be decoded as the corresponding keys.
To achieve this goal, we first use our decoder $D$ to replace the classifier $h_s$ in the suspect model $M_s$, then the suspect model can be restructured as $M_s=(f_s,D)$.
According to \cref{def:wm_coll}, we test the suspect model $M_s$ on all the verification datasets $\cD_T=\{\cD_{T_1}, \dots, \cD_{T_k}, \dots, \cD_{T_K}\}$ 
of all the clients
to track the malicious clients, and report $\text{WSR}_k$ on the $k$-th verification datasets correspondingly. 
The client whose verification
dataset achieves the highest WSR leaked the model (see \cref{fig:framework} server verification). The tracking mechanism can be defined as $\textit{Track}(M_s, \cD_T) = \argmax_{k} \text{WSR}_k$.

Suppose the ground truth malicious client is $k_m$. 
If $\text{WSR}_{k_m}>\sigma$, and  $\text{WSR}_{k}$ for other verification datasets is smaller than $\sigma$, then the ownership of the model can be verified, and no watermark collision happens. If  $\textit{Track}(M_s, \cD_T)=k_m$, then the malicious client is identified correctly.

\begin{algorithm}[htbp]
   \caption{Injection of Decodable Unique Watermarking (DUW)}
   \label{algorithm:inject}
\begin{algorithmic}[1]
   \STATE {\algorithmicrequire} %
   Clients datasets\{$\cD_k\}_{k=1}^K$,
   OoD dataset  $\cD_{\text{OoD}}$,
  secret key $\{s_k\}_{k=1}^K$,
  pre-trained encoder $E$,
   pre-defined decoder $D$,
   global parameters $\theta_g$, local parameters $\{\theta_k\}_{k=1}^K$,
            learning rate $\alpha$,$\beta$, local training steps $T$, watermark injection steps $T_w$.
\\ \STATE{\textbf{Step 1: Client-unique trigger encoding.}}
 \FOR{$k=1$,$\dots,K$}
  \STATE{Generate trigger set for client $k$: $\cD_{T_k}=\{(x',s_k)|x' \sim E_{x\in \cD_{OoD}}(x, s_k;\theta_E)\}$}
 \ENDFOR
 \\ \STATE{\textbf{Step 2: Decoding triggers to client keys.}}
     \REPEAT    
    \STATE{Server selects active clients $\cA$ uniformly at random}
    
   \FOR{all client $k \in \cA$}
       \STATE{Server initializes watermarked model for client $k$ as: $\theta_k \leftarrow \theta_g$.}

       \FOR{$t=1$, $\dots, T_w$}  

           \STATE{Server replaces model classifier $h$ with decoder $D$.}
           \STATE{
           Server injects watermark to model using trigger set $\cD_{T_k}$, and update $\theta^f_k$ as:\\
           $\theta^f_k \leftarrow \theta^f_k - \beta \nabla_{\theta^f_k} J'(\theta^f_k).$ \hspace{0.1in} $\triangleright$ Optimize \cref{eq:l2} }
       \ENDFOR
        \STATE{Server broadcasts $\theta_k$ to the corresponding client $k$.}
 \FOR{$t=1$, $\dots, T$}  
\STATE{Client local training using local set $\cD_k$: $\theta_k \leftarrow \theta_k - \alpha \nabla_{\theta_k} J_{k}(\theta_k)$. \hspace{0.1in} $\triangleright$ Optimize \cref{eq:local_train} }
    \ENDFOR    
       \STATE{Client $k$ sends $\vtheta_k$ back to the server.}
\ENDFOR
     \STATE{Server updates $\theta_g \leftarrow \frac{1}{|\cA|}\sum_{k \in \cA} \theta_k$.} 
\UNTIL{training stop}
\end{algorithmic}
\end{algorithm}

%% file: sec/exp.tex
\section{Experiments}
\label{sec:exp}
In this section, we empirically show how our proposed DUW can fulfill the requirements (\underline{R1-R4}) for tracking infringers as described in \cref{sec:intro}.\\
\textbf{Datasets.}
To simulate \emph{class non-iid} FL setting, we use CIFAR-10, CIFAR-100~\citep{krizhevsky2009learning}, which contain $32 \times 32$ images with $10$ and $100$ classes, respectively. CIFAR-10 data is uniformly split into $100$ clients, and $3$ random classes are assigned to each client. CIFAR-100 data is split into $100$ clients with Dirichlet distribution.
For CIFAR-10 and CIFAR-100, the OoD dataset we used for OoD injection is a subset of ImageNet-DS~\citep{chrabaszcz2017downsampled} with randomly chosen $500$ samples downsampled to $32\times32$.
To simulate the \emph{feature non-iid} FL setting, a multi-domain FL benchmark, Digits~\citep{li2020federated,hong2022efficient} is adopted. The dataset is composed of $28 \times 28$ images for recognizing $10$ digit classes, which was widely used in the community~\citep{caldas2018leaf,mcmahan2017communication}. The Digits includes five different domains: MNIST~\citep{lecun1998gradient}, SVHN~\citep{netzer2011reading}, USPS~\citep{hull1994database}, SynthDigits~\citep{ganin2015unsupervised}, and MNIST-M~\citep{ganin2015unsupervised}. We leave out USPS as the OoD dataset for watermark injection ($500$ samples are chosen) and use the rest four domains for the standard FL training. Each domain of digits is split into $10$ different clients, thus, $40$ clients will participate in the FL training.
\\\textbf{Training setup.}
A preactivated ResNet (PreResNet18)~\citep{he2016deep} is used for CIFAR-10, a preactivated ResNet
(PreResNet50)~\citep{he2016deep} is used for CIFAR-100, and a CNN defined in~\cite{li2021fedbn} is used for Digits.
For all three datasets, we leave out $10\%$ of the training set as the validation dataset to select the best FL model.
The total training round is $300$ for CIFAR-10 and CIFAR-100, and $150$ for Digits.
\\\textbf{Watermark injection.}
The early training stage of FL is not worth protecting since the standard accuracy is very low, we start watermark injection at round $20$ for CIFAR-10 and Digits, and at round $40$ for CIFAR-100. The standard accuracy before our watermark injection is $85.20\%$, $40.23\%$, and $29.41\%$  for Digits, CIFAR-10, and CIFAR-100, respectively.
\\\textbf{Evaluation metrics.}
For watermark verification, we use watermark success rate (\textbf{WSR}) which is the accuracy of 
the trigger set
for evaluation. To measure whether we track the malicious client (leaker) correctly,  we define tracking accuracy (\textbf{TAcc}) as the rate of the clients we track correctly. To further evaluate the ability of our method for distinguishing between different watermarks for different clients, we also report the difference between the highest WSR and second best WSR as \textbf{WSR\_Gap} to show the significance of verification and IP tracking. With a significant WSR\_Gap, no watermark collision will happen. 
To evaluate the utility of the model, we report the standard FL accuracy (\textbf{Acc}) for each client’s individual test sets, whose classes match their training sets. We also report the accuracy degradation (\textbf{$\Delta$Acc}) of the watermarked model compared with the non-watermarked one.
Note that, 
to simulate the scenario where malicious clients leak their local model after local training, 
we test the average WSR, TAcc and WSR\_Gap for the local model of each client instead of the global model.
Acc and $\Delta$Acc are evaluated on the best FL model selected using the validation datasets.

\subsection{IP Tracking Benchmark}
We evaluate our method using the IP tracking benchmark with various metrics as shown in \cref{tab:benchmark_results}. Our ownership verification is confident with
all WSRs over $99\%$ (\underline{R2}). The model utility is also preserved with accuracy degradation $2.34\%$, $0.03\%$, and $0.63\%$, respectively for Digits, CIFAR-10 and CIFAR-100  (\underline{R3}).  TAcc for all benchmark datasets is $100\%$ which indicates accurate IP tracking (\underline{R1}).
All WSR\_Gap is over $98\%$, which means the WSRs for all other benign client's verification datasets are close to $0\%$. In this way, the malicious client can be tracked accurately with high confidence, no collisions will occur within our tracking mechanism (\underline{R1}). 

\begin{table}[t]
    \small
    \centering
    \begin{tabular}{cccccc}
    \toprule
         Dataset&Acc&$\Delta$Acc&WSR&WSR\_Gap&TAcc  \\
         \hline
        Digits&0.8855&0.0234&0.9909&0.9895&1.0000\\
         CIFAR-10&0.5583 &0.0003 &1.0000&0.9998&1.0000\\
         CIFAR-100&0.5745&0.0063&1.0000&0.9998&1.0000\\
         \bottomrule
    \end{tabular}
    \caption{Benchmark results.}
    \label{tab:benchmark_results}
    \vspace{-0.3in}
\end{table}

\subsection{Robustness}
\label{sec:robustness}
Malicious clients can conduct watermark removal attacks before leaking the FL model to make it harder for us to verify the model copyright, and track the IP infringers accurately.
In this section, we show the robustness of the watermarks under various watermark removal attacks (\underline{R4}).
Specifically, we evaluate our method against 1) \textbf{fine-tuning}~\citep{adi2018turning}:
Fine-tune the model using their own local data; 2) \textbf{pruning}~\citep{liu2018rethinking}:
prune the model parameters that have the smallest absolute value according to a certain pruning rate, and then fine-tune the model on their local data; 
3) \textbf{model extraction attack}: first query the victim model for the label of an auxiliary dataset, and then re-train the victim model on the annotated dataset.
We take knockoff~\citep{orekondy2019knockoff} as an example of the model extraction attack;
4) \textbf{parameter perturbations}: add random noise to local model parameters~\citep{garg2020can}. 

$10$ of the clients are selected as the malicious clients, and the metrics in this section are average values for $10$ malicious clients. All the watermark removal attacks are conducted for $50$ epochs with a learning rate $10^{-5}$. All the attacks are conducted for the local model of the last round.

\textbf{Robustness against fine-tuning attack.}
We report the robustness of our proposed DUW against fine-tuning in \cref{tab:finetune}.  $\Delta$Acc and $\Delta$WSR in this table indicate the accuracy and WSR drop compared with accuracy and WSR before the attack. According to the results, after $50$ epochs of fine-tuning, the attacker can only decrease the WSR by less than $1\%$, and the TAcc is even not affected. Fine-tuning with their limited local training samples can also cause a standard accuracy degradation. 
Fine-tuning can neither remove our watermark nor affect our IP tracking, even if sacrifices their standard accuracy.

\begin{table}[t]
    \small
    \centering
    \begin{minipage}{0.485\textwidth} 
    \setlength{\tabcolsep}{1.0mm}{
    \scalebox{0.9}{
    \begin{tabular}{ccccccc}
    \toprule
         Dataset&Acc&$\Delta$Acc&WSR&$\Delta$WSR&TAcc  \\
         \hline
        Digits&0.9712&-0.0258&0.9924&0.0030&1.0000\\
         CIFAR-10&0.7933&0.1521&1.0000&0.0000&1.0000\\
         CIFAR-100&0.4580&0.0290&0.9930&0.0070&1.0000\\
         \bottomrule
    \end{tabular}}}
    \caption{DUW is robust against fine-tuning.}
    \label{tab:finetune}
    \end{minipage}
    \begin{minipage}{0.485\textwidth} \setlength{\tabcolsep}{1mm}{
    \scalebox{0.9}{
    \begin{tabular}{ccccccc}
    \toprule
         Dataset&Acc&$\Delta$Acc&WSR&$\Delta$WSR&TAcc  \\
         \hline
        Digits&0.8811&0.0643&0.9780&0.0174&1.0000\\
         CIFAR-10&0.5176&0.0010&0.6638&0.3362&1.0000\\
         CIFAR-100&0.4190&0.0680&0.8828&0.1172&1.0000\\
         \bottomrule
    \end{tabular}}}
    \caption{DUW is robust against model extraction. 
    }
    \label{tab:extraction}
    \end{minipage}
    \vspace{-0.2in}
\end{table}

\begin{figure}
    \centering
    \begin{subfigure}  {0.32\textwidth}\centerline{\includegraphics[width=4.6cm]{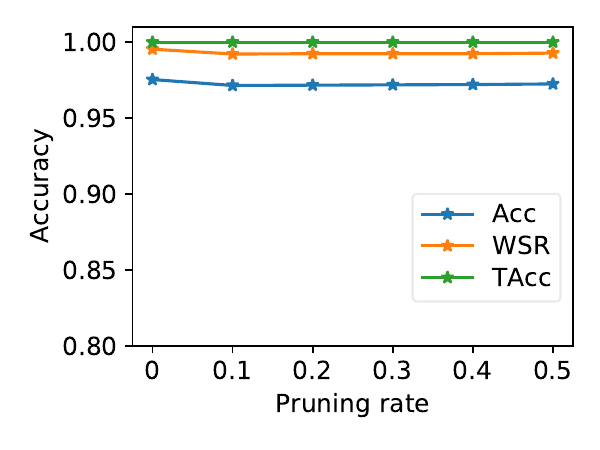}}
        \vspace{-0.1in}
        \subcaption{Digits.}
    \end{subfigure}
     \begin{subfigure}  {0.32\textwidth}\centerline{\includegraphics[width=4.6cm]{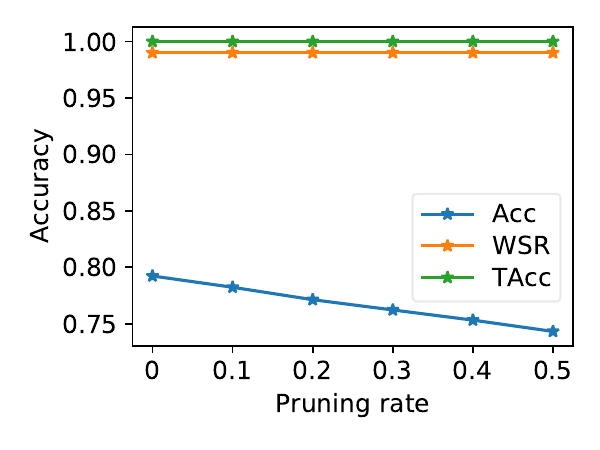}}
     \vspace{-0.1in}
        \subcaption{CIFAR-10.}
    \end{subfigure}
     \begin{subfigure}  {0.32\textwidth}\centerline{\includegraphics[width=4.6cm]{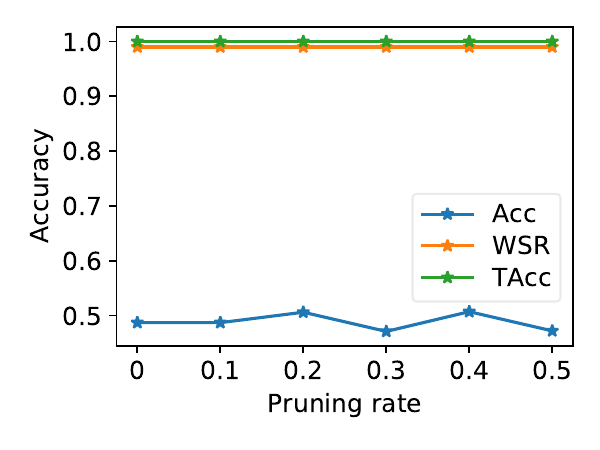}}
     \vspace{-0.1in}
        \subcaption{CIFAR-100.}
    \end{subfigure}
    \caption{DUW is robust against pruning.}
    \label{fig:prune}
    \vspace{-0.3in}
\end{figure}
\textbf{Robustness against pruning attack.}
We investigate the effect of pruning in \cref{fig:prune}
by varying the pruning rate from $0$ to $0.5$. With the increase in the pruning ratio, both TAcc and WSR will not be affected. For CIFAR-10, standard accuracy will drop $5\%$. Therefore, pruning is not an effective attack on our watermark, and it will even cause an accuracy degradation for the classification task.

\textbf{Robustness against model extraction attack.}
To verify the robustness of our proposed DUW against model extraction attack, we take knockoff~\citep{orekondy2019knockoff} as an example, and STL10~\citep{coates2011analysis} cropped to the same size as the training data is used as the auxiliary dataset for this attack. According to the results for three benchmark datasets in \cref{tab:extraction}, after knockoff attack, WSR for all three datasets is still over $65\%$, and our tracking mechanism is still not affected with TAcc remains to be $100\%$. Therefore, our DUW is resilient to model extraction attacks.

\begin{figure}[htbp!]
    \centering
    \vspace{-0.2in}
    \begin{subfigure}  {0.32\textwidth}\centerline{\includegraphics[width=4.6cm]{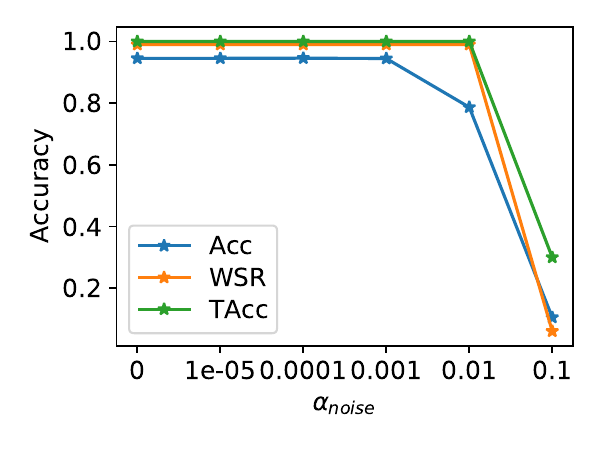}}
    \vspace{-0.15in}
        \subcaption{Digits.}
    \end{subfigure}
     \begin{subfigure}  {0.32\textwidth}\centerline{\includegraphics[width=4.6cm]{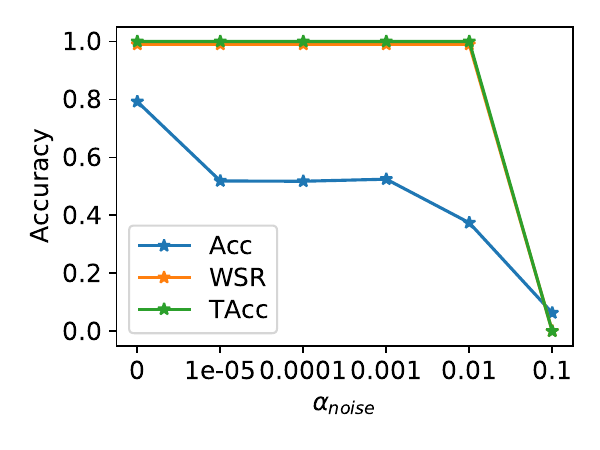}}
     \vspace{-0.15in}
        \subcaption{CIFAR-10.}
    \end{subfigure}
     \begin{subfigure}  {0.32\textwidth}\centerline{\includegraphics[width=4.6cm]{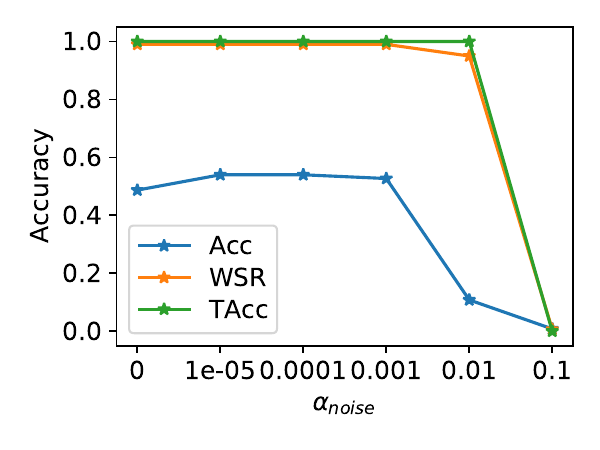}}
     \vspace{-0.15in}
        \subcaption{CIFAR-100.}
    \end{subfigure}
    \caption{DUW is robust against parameter perturbation.
    }
    \vspace{-0.2in}
    \label{fig:noise}
\end{figure}

\textbf{Robustness against parameter perturbations attack.}
Malicious clients can also add random noise to model parameters to remove watermarks, since \cite{garg2020can} found that backdoor-based watermarks are usually not resilient to parameter perturbations. Adding random noise to the local model parameters can also increase the chance of blurring the difference between different watermarked models. We enable each malicious client to blend Gaussian noise to the parameters of their local model, and set the parameter of the local model as $\theta_i=\theta_i+\theta_i*\alpha_{\text{noise}}$, where $\alpha_{\text{noise}}=\{10^{-5},10^{-4}, 10^{-3},10^{-2},10^{-1}\}$.
We investigate the effect of parameter perturbation in \cref{fig:noise}. According to the results, when $\alpha_{\text{noise}}$ is smaller than $10^{-2}$, WSR, Acc, and TAcc will not be affected. When $\alpha_{\text{noise}}=10^{-2}$, Acc will drop more than $10\%$, TAcc remains unchanged, and WSR is still over $90\%$. When $\alpha_{\text{noise}}=10^{-1}$, Acc will drop to a random guess, thus, although the watermark has been removed, the model has no utility.
Therefore, parameter perturbation is not an effective attack for removing our watermark and affecting our tracking mechanism.

\subsection{Qualitative Study}

\textbf{Effects of decoder.}
To investigate the effects of the decoder on avoiding watermark collision, we compare the results of w/ and w/o decoder. When the decoder is removed, the task dimension of the watermark injection will be the same as the FL classification, thus, we also have to change the original target label (the same as the input key)  to the FL classification task dimension. To achieve this goal, we set the target label of w/o decoder case as (client\_ID $\%$ class\_number). 
We report the results of w/ and w/o decoder on CIFAR-10 after $1$ round of watermark injection at round $20$ in \cref{tab:decoder}. 
According to the results, when we have 100 clients in total, w/o decoder can only achieve a TAcc of $6\%$, while w/ decoder can increase TAcc to $100\%$. We also find that clients with the same target label are more likely to conflict with each other, which makes those clients difficult to be identified, even if their trigger sets are different. Utilizing a decoder to increase the target label space to a dimension larger than the client number allows all the clients to have their own target label. In this way, watermark collision can be avoided. Besides, WSR of w/ decoder is also higher than w/o decoder after $1$ round of injection. One possible reason is that we differ the watermark injection task from the original classification task using the decoder, thus, in this case, the watermark will be more easily injected compared with directly injected to the original FL classification task.

\begin{figure}[!ht]
\begin{center}
\vspace{-0.1in}
\begin{subfigure}{0.49\linewidth}
\includegraphics[width=3.3cm]{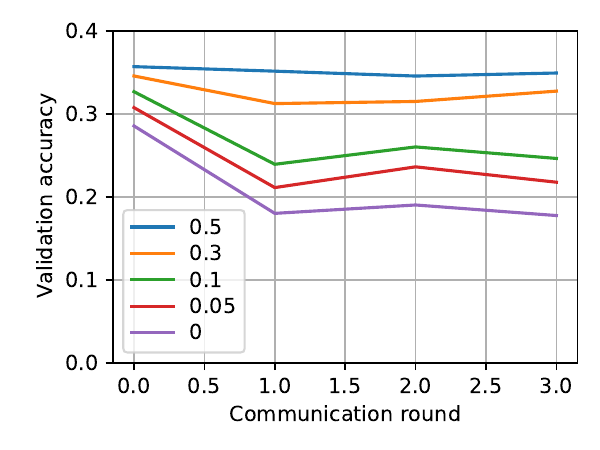}
\includegraphics[width=3.3cm]{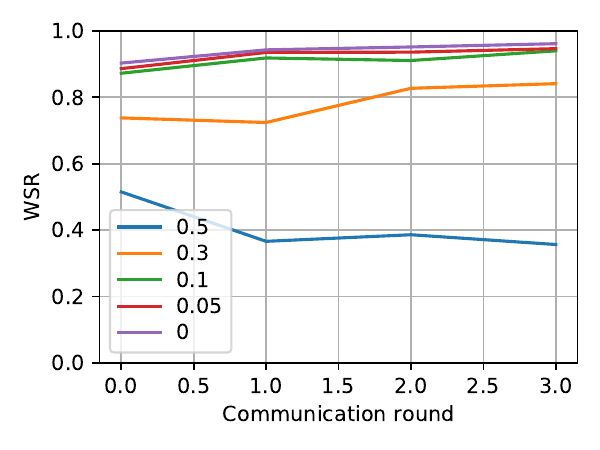}
 \vspace{-0.05in}
\caption{Acc and WSR for different values of $\beta$.}
\label{fig:beta_iter}
\end{subfigure}
\begin{subfigure}{0.49\linewidth}
\includegraphics[width=3.3cm]{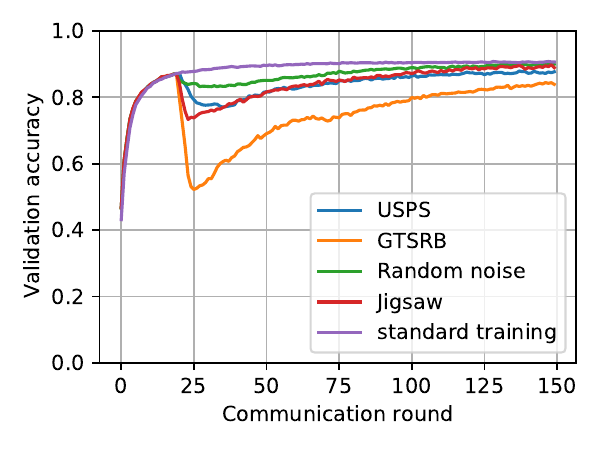}
\includegraphics[width=3.3cm]{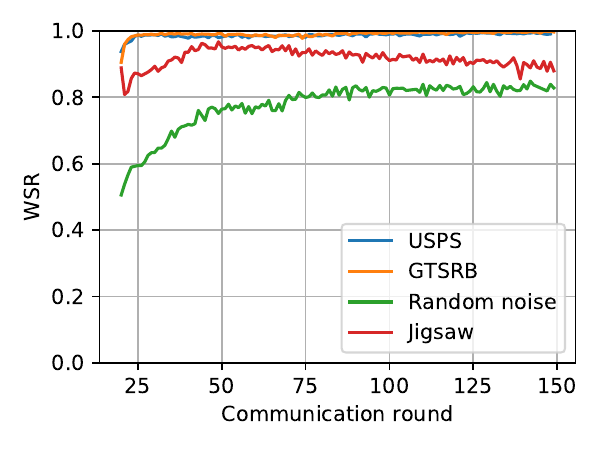}
 \vspace{-0.05in}
\caption{Acc and WSR for different OoD datasets.}
 \label{fig:ood_iter}
 \end{subfigure}
 \vspace{-0.1in}
 \caption{Acc and WSR w.r.t. different communication rounds.}
 \vspace{-0.2in}
\end{center}
\end{figure}

\textbf{Effects of $l_2$ regularization.}
To show the effects of $l_2$ regularization in \cref{eq:l2}, we report the validation accuracy and WSR for $4$ rounds of watermark injection on Digits with different values of the hyperparameter $\beta$
in \cref{fig:beta_iter}. 
Validation accuracy is the standard FL accuracy evaluated on a validation dataset for every round.
We see that with the increase of $\beta$, higher validation accuracy can be achieved, but correspondingly, WSR drops from over $90\%$ to only $35.65\%$. 
Larger $\beta$ increases the impact of $l_2$ norm, which decreases the model difference between the watermarked model and the non-watermarked one, so the validation accuracy will increase. 
At the same time, the updates during watermark injection also have much more restriction due to $l_2$ regularization, so the WSR drops to a low value. 
Accordingly, we select $\beta=0.1$ for all our experiments, since $\beta=0.1$ can increase validation accuracy by $6.88\%$ compared with $\beta=0$, while maintaining WSR over $90\%$.

\textbf{Effects of different OoD datasets for watermark injection.}
We investigate the effects of different OoD datasets including USPS~\citep{hull1994database}, GTSRB~\citep{stallkamp2012man}, random noise, and Jigsaw  for watermark injection when the standard training data is Digits. 
All OoD images are cropped to the same size as the training images. 
A jigsaw image is generated from a small $4\times4$ random image, and then uses reflect padding mode from PyTorch to padding to the same size as the training images. 
The effect of these different OoD datasets is shown in \cref{tab:ooddataset} and \cref{fig:ood_iter}. 
We see that all OoD datasets can achieve $100\%$ TAcc, suggesting the selection of OoD dataset will not affect the tracking of the malicious client.
There is a trade-off between the Acc and WSR: higher WSR always leads to lower Acc. 
Random noise and jigsaw achieve high Acc, with accuracy degradation within $1\%$. 
These two noise OoD also have a faster recovery of the standard accuracy after the accuracy drop at the watermark injection round as shown in \cref{fig:ood_iter}, but the WSR of random noise and Jigsaw are lower than $90\%$. For two real OoD datasets USPS and GTSRB, the WSR quickly reaches over $99\%$ after $1$ communication round, but their accuracy degradation is larger than $2\%$. 

\begin{table}[t]
    \centering
    \small
    \begin{minipage}{0.45\textwidth} \setlength{\tabcolsep}{1.2mm}{\scalebox{1.0}{
        \begin{tabular}{ccccc}
    \toprule
         Method&Acc&$\Delta$Acc&WSR&TAcc  \\
         \hline
         w/ decoder&0.3287&0.0736&\textbf{0.8778}&\textbf{1.0000}\\
w/o decoder&0.3235&0.0788&0.8099&0.0600\\
         \bottomrule
    \end{tabular}}}
    \caption{Effects of decoder: the decoder can improve TAcc to avoid watermark collision. $\Delta$Acc in this table is the accuracy degradation compared with the previous round.}
    \label{tab:decoder}
    \end{minipage}\hspace{+0.1in}
    \begin{minipage}{0.485\textwidth} \setlength{\tabcolsep}{1.0mm}{\scalebox{0.9}{
            \begin{tabular}{cccccc}
    \toprule
         Dataset&Acc&$\Delta$Acc&WSR&WSR\_Gap&TAcc  \\
         \hline
         USPS&
    0.8855&0.0234&0.9909&0.9895&1.0000\\
         GTSRB&0.8716& 0.0373 &0.9972&0.9962&1.0000\\
         Random noise&0.9007&0.0082 &0.8422&0.8143&1.0000\\
         Jigsaw&0.9013&0.0076 &0.8789&0.8601&1.0000\\
         \bottomrule
    \end{tabular}}}
    \caption{Effects of different OoD datasets: a trade-off exists between Acc and WSR, given different selections of OoD datasets.}
    \label{tab:ooddataset}
    \end{minipage}
    \vspace{-0.3in}
\end{table}

\textbf{Scalability of DUW to more clients.}
We conduct an ablation study to show the effect of the number of clients in \cref{tab:diff_client}. According to the results, even with $600$ clients, the WSR is still over $73\%$ and the TAcc remains $100\%$. With more clients participating in FL, we can still track the malicious client correctly with high confidence.
\begin{table}[htbp!]
    \small
    \centering
    \begin{tabular}{cccccc}
    \toprule
         Number of clients&Acc&$\Delta$Acc&WSR&WSR\_Gap&TAcc  \\
         \hline
40&0.8855&0.0234&0.9909&0.9895&1.0000\\
400&0.8597&-0.0332&0.9521&0.9267&1.0000\\
600&0.8276&-0.0035&0.7337&0.6383&1.0000\\
         \bottomrule
    \end{tabular}
    \caption{Ablation study: results for different numbers of clients on digits. }
    \label{tab:diff_client}
\end{table}

\textbf{Hybrid watermark.}
If DUW meets a black-box suspect model,
 our DUW can also be combined with existing black-box unified watermarks. We can identify IP leakage using black-box detection with a unified watermark first, then identify infringers using DUW with client-unique watermark.
 We design a simple hybrid watermark in this section as an example. We pick one of the trigger sets we generated for the clients as the trigger set for the unified watermark injection, and the target label is assigned as $0$ which belongs to the original label set of the training data. We use this trigger set to fine-tune the entire global model for $10$ steps before injecting our proposed DUW. Note that no decoder is used for the unified watermark, and the unified watermarks can also be replaced with other existing works. The results on Digits are shown in \cref{tab:hybrid_watermark}. For this table, we can observe that the unified watermark is injected successfully in the presence of our DUW, with a $98.82\%$ WSR. Besides, the effectiveness of our DUW is also not affected,
since the WSR of DUW only decreases by $0.72\%$, and TAcc remains $100\%$. The model utility is also not affected, since the standard accuracy remains high.
\begin{table}[htbp!]
    \small
    \centering
    \begin{tabular}{ccccccc}
    \toprule
         Method&Acc&$\Delta$Acc&WSR&WSR\_Gap&TAcc&Unified WSR  \\
         \hline
        w/o unified watermark&0.8855&0.0234&0.9909&0.9895&1.0000&/ \\
       w/ unified watermark&0.8886&0.0203&0.9837&0.9701&1.0000&0.9882\\
         \bottomrule
    \end{tabular}
    \caption{Results for hybrid watermark.}
    \label{tab:hybrid_watermark}
    \vspace{-0.25in}
\end{table}

%% file: sec/appendix.tex
\section{Discussions}
\textbf{Client-side watermarking VS server-side watermarking.}
Client-side watermarking such as FedCIP~\citep{liang2023fedcip}, and Merkle-Sign~\citep{li2021towards} are used to claim the co-ownership of the model, yet we argue that client-side watermarking has some limitations, which makes it not applicable for IP tracking. For client-side watermarking, if one of the clients is the infringer to illegally distribute the model, the infringer will not reveal their own identity during the model verification process in order to avoid legal responsibility. Even if the ownership of the model can be claimed by their co-author, the real infringers cannot be tracked, since they remain anonymous. Using our server-side watermark, there is no such concern, the server can easily track the malicious client among all the clients. 

\textbf{Complexity.}
Clients will not experience additional computations as our DUW is carried out on the server side. The additional computation for the server is decided by the number of watermark injection steps $T_w$. We found that WSR could reach $99\%$ just within $T_w=10$ steps. Injection of one client-unique watermark takes around $1$ second. The server can embed the watermark parallelly for all the clients. Since the watermarked model for each client is independent and has no sequence relationship with each other, there is no need to serialize it.
Thus, the delay caused by the server is neglectable.

\textbf{Future works.}
This paper makes the FL model leakage from anonymity to accountability by injecting client-unique watermarks. 
We recognize the most significant challenge for accountable FL is addressing watermark collision for accurate IP tracking (\underline{R1}). 
We believe it is important to scale our method with more clients in the future. One plausible solution is increasing the dimension of the input of the encoder to allow more one-hot encoding target labels. Another solution is to use a hash function as the target label for different clients. In this way, the lower-dimensional encoder and decoder can accommodate more clients. For instance, an encoder with input dimension $10$ can allow at most $1024$ different clients. However, adopting hash functions as the target labels can increase the chance of watermark collision between clients, and more elegant strategies have to be developed to address this problem.
As we focus on the collision, we leave the scalability for future work.

\section{Supplementary experiments}
\subsection{Comparison with traditional backdoor-based watermarks}\label{sec:baseline}
\begin{figure}[htbp!]
    \centering
    \begin{subfigure}  {0.32\textwidth}\centerline{\includegraphics[width=4.6cm]{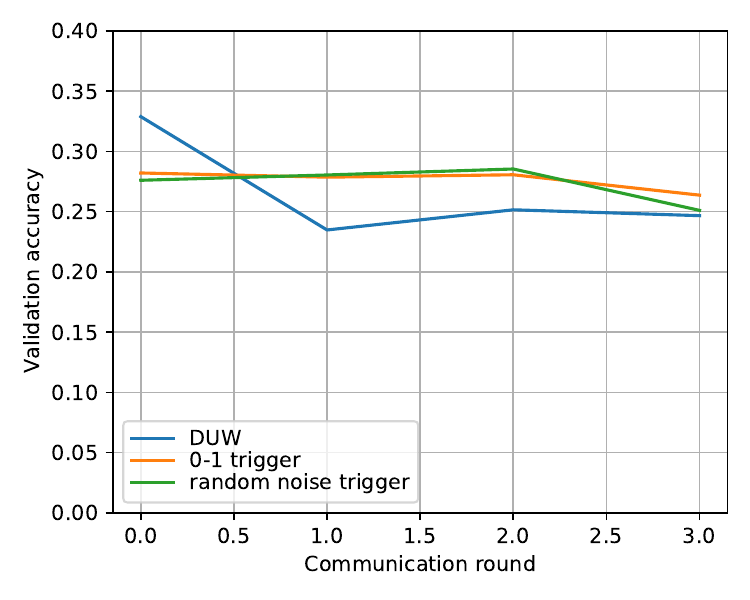}}
    \vspace{-0.15in}
        \subcaption{Validation accuracy.}
    \end{subfigure}
     \begin{subfigure}  {0.32\textwidth}\centerline{\includegraphics[width=4.6cm]{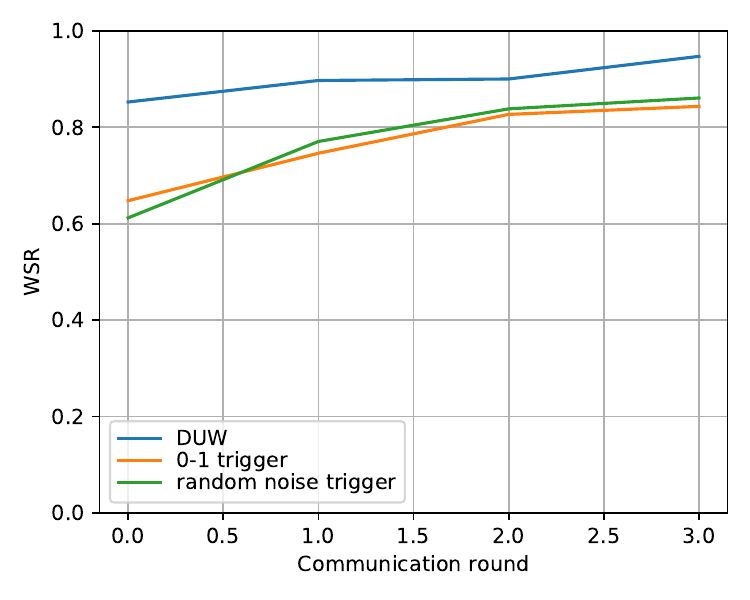}}
     \vspace{-0.15in}
        \subcaption{WSR.}
    \end{subfigure}
     \begin{subfigure}  {0.32\textwidth}\centerline{\includegraphics[width=4.6cm]{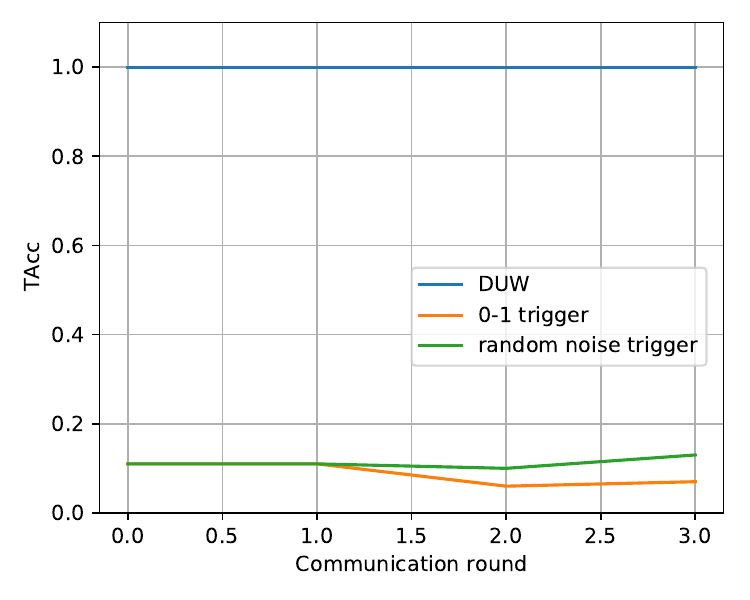}}
     \vspace{-0.15in}
        \subcaption{TAcc.}
    \end{subfigure}
    \caption{Validation accuracy, WSR, and TAcc for proposed DUW and other two baselines on CIFAR-10 for $4$ communication rounds. 
    }
    \vspace{-0.15in}
    \label{fig:diff_trigger}
\end{figure}

We compare our proposed DUW with two traditional backdoor-based watermarks in \cref{fig:diff_trigger}. Due to the reason that if all the clients share the same trigger, watermark collision will definitely happen, we design different triggers for different clients. Specifically, we use traditional backdoor-based watermarking by adding arbitrary badnet triggers using random noise or 0-1 coding trigger for each client. To distinguish between different clients, for 0-1 trigger, following ~\cite{tang2020embarrassingly}, we set $5$ pixel values of the pattern into $0$ and other $11$ pixels into 1, different combinations of the pattern are randomly chosen for different clients. For random noise triggers, we generate different random noise triggers for different clients.
The trigger size $4\times4$ and the injection is conducted for $4$ rounds. The target label for each client is set as (client\_{ID} \% class\_number). According to the results, traditional backdoor-based watermarks can only achieve a tracking accuracy lower than $13\%$ (it will even be lower with the increase of the communication rounds), which is much lower than the $100\%$ tracking accuracy we have achieved. Note that, the rate of clients with watermark collisions can be calculated as 1-TAcc.

To analyze the failure of the traditional backdoor-based watermarking, 
we give detailed prediction results for one trial on CIFAR10 for random noise trigger as an example. The client number is $100$, so the client\_{ID} is from 0-99, and the class number is $10$. Here we provide a fine-grained analysis of the concerned $13\%$ TAcc by looking into the last $10$ clients. We list the client\_{ID} and their corresponding predicted client\_{ID} for clients 90-99 in \cref{tab:detail_pred}. From the prediction results, $8$ of $10$ clients are tracked wrong. Among these $8$ failure cases, $7$ of the predicted client\_{ID} (client $90$, $91$, $93$, $94$ $95$, $96$, $98$)  share the same targets, and 1 of them (client 97) have both different triggers and different target labels. The two kinds of failures correspond to two different reasons respectively as we illustrated in \cref{sec:pitfall}. 
\begin{table}[htbp!]
    \centering
    \begin{tabular}{c|c}
    \toprule
ground truth client\_{ID}&	predicted client\_{ID}  \\
\hline
    90&	0\\
91&	1\\
92&	92\\
93&	3\\
94&	4\\
95&	5\\
96&	6\\
97&	49\\
98&	8\\
99&	99\\  
\bottomrule
    \end{tabular}
    \caption{Prediction results for random noise trigger for client 90-99.}
    \label{tab:detail_pred}
\end{table}

\subsection{Extended qualitative study}

\begin{figure}[htbp!]
    \centering
    \begin{subfigure}{0.23\textwidth}\centerline{\includegraphics[width=2cm]{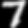}}
        \subcaption{Digits: original.}
        \end{subfigure}  
        \begin{subfigure}{0.24\textwidth}
            \centerline{\includegraphics[width=2cm]{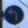}}
        \subcaption{GTSRB: original.}
        
        \end{subfigure}  
          \begin{subfigure} {0.23\textwidth}
            \centerline{\includegraphics[width=2cm]{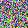}}
        \subcaption{Noise: original. }
        \end{subfigure}  
    \begin{subfigure}{0.23\textwidth}
            \centerline{\includegraphics[width=2cm]{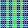}}
        \subcaption{Jigsaw: originaL.}
        \end{subfigure}  
              \begin{subfigure} {0.23\textwidth}
            \centerline{\includegraphics[width=2cm]{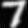}}
        \subcaption{Digits: encoded.}
        \end{subfigure}  
         \begin{subfigure}{0.23\textwidth}
            \centerline{\includegraphics[width=2cm]{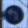}}
        \subcaption{GTSRB: encoded.}
        \end{subfigure}
         \begin{subfigure} {0.23\textwidth}
            \centerline{\includegraphics[width=2cm]{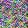}}
        \subcaption{Noise: encoded. }
        \end{subfigure}  
        \begin{subfigure}{0.23\textwidth}
            \centerline{\includegraphics[width=2cm]{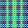}}
        \subcaption{Jigsaw: encoded.}
        \end{subfigure}  
              \begin{subfigure} {0.23\textwidth}
            \centerline{\includegraphics[width=2cm]{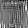}}
        \subcaption{Digits: residual.}
        \end{subfigure}  
         \begin{subfigure}{0.23\textwidth}
            \centerline{\includegraphics[width=2cm]{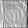}}
        \subcaption{GTSRB: residual.}
        \end{subfigure}
         \begin{subfigure} {0.23\textwidth}
            \centerline{\includegraphics[width=2cm]{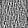}}
        \subcaption{Noise: residual. }
        \end{subfigure}  
        \begin{subfigure}{0.23\textwidth}
            \centerline{\includegraphics[width=2cm]{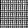}}
        \subcaption{Jigsaw: residual.}
        \end{subfigure}  
    \caption{Visualization of unique trigger set based on different OoD datasets.
    }
    \label{fig:vis_ood}
\end{figure}

\textbf{Visualization of unique trigger sets.}
We show the visualization example for the original image, encoded image (image in trigger set), and residual image based on different OoD datasets in \cref{fig:vis_ood}. We observe that for all four different OoD datasets, the original image and encoded image with our client keys are indistinguishable from the human eye. The difference between these two images can be observed in the residual image. Note that although the OoD datasets are different, the encoder that we used to generate the trigger sets is the same. According to \cref{fig:vis_ood}, the encoder will generate sample-wise triggers for different images.

\begin{figure}[htbp!]
    \centering
    \begin{subfigure}{0.18\textwidth}\centerline{\includegraphics[width=2cm]{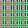}}
        \subcaption{Original image example.}
        \end{subfigure}  
        \begin{subfigure}{0.18\textwidth}
            \centerline{\includegraphics[width=2cm]{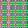}}
        \subcaption{Encoded image for client 0.}
        
        \end{subfigure}  
            \begin{subfigure}{0.18\textwidth}
            \centerline{\includegraphics[width=2cm]{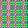}}
        \subcaption{Encoded image for client 1.}
        \end{subfigure}  
          \begin{subfigure} {0.18\textwidth}
            \centerline{\includegraphics[width=2cm]{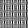}}
        \subcaption{Residual image for client 0. }
        \end{subfigure}  
              \begin{subfigure} {0.18\textwidth}
            \centerline{\includegraphics[width=2cm]{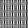}}
        \subcaption{Residual image for client 1.}
        \end{subfigure}  
    \caption{Visualization of the unique trigger sets for two different clients. The difference between trigger sets cannot be observed according to human inspection, but after decoding, the difference between keys can be distinguished by our model.
    }
    \label{fig:vis_trigger}
\end{figure}

To investigate the difference between different clients' trigger sets based on the same OoD dataset, we show one example in the trigger set generated by the jigsaw image for two randomly picked clients in \cref{fig:vis_trigger}. The trigger sets are generated based on the same jigsaw dataset and differ by their embedded keys. According to \cref{fig:vis_trigger}, 
although the samples from different trigger sets do not look distinguishable according
to the human inspection, the difference between keys decoded from the trigger sets can be distinguished by our model.

\begin{figure}[htbp!]
    \centering
    \begin{subfigure}  {0.45\textwidth}\centerline{\includegraphics[width=5cm]{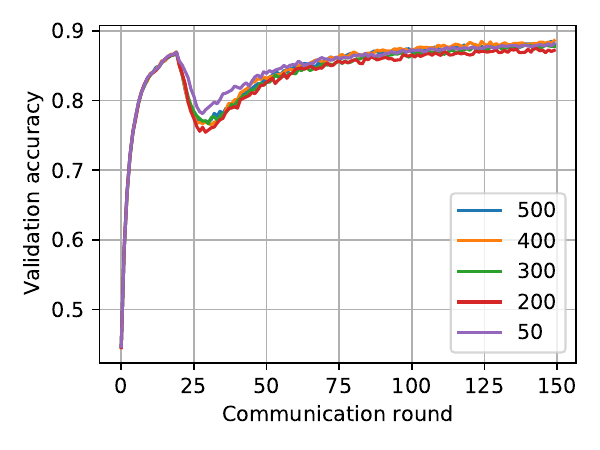}}
    \vspace{-0.15in}
        \subcaption{Validation accuracy.}
    \end{subfigure}
     \begin{subfigure}  {0.45\textwidth}\centerline{\includegraphics[width=5cm]{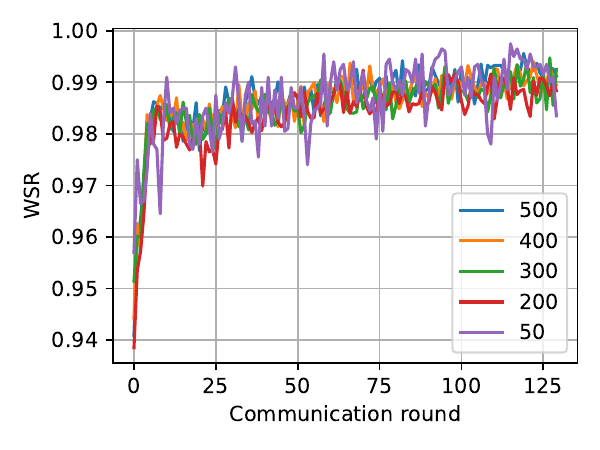}}
     \vspace{-0.15in}
        \subcaption{WSR.}
    \end{subfigure}
    \caption{Effects of the different number of samples in trigger sets. $50$ samples in one trigger set can achieve over $98\%$ WSR.
    }
    \label{fig:diff_num}
\end{figure}
\textbf{Effects of the different numbers of samples in trigger sets.}
We investigate how the size of the trigger set will affect our watermark injection and standard FL training in \cref{fig:diff_num} by varying the number of samples in the trigger set from $50$ to $500$ for Digits training (USPS is used to generate the trigger set). Note that for all cases, TAcc always remains to be $100\%$.  We observe that with only $50$ samples in one trigger set, we can achieve an accuracy degradation around $2\%$, and with a WSR over $98\%$. When the number of samples increases to $300$, WSR is over $99\%$. In general, the change in the number of samples in the trigger set has almost no effect on both standard accuracy and WSR. A small trigger set (such as $50$) can achieve comparable results with a large trigger set. 
The advantages of a smaller trigger set include quicker trigger set generation, quicker watermark injection, quicker ownership verification, and quicker IP tracking. Besides, less effort can be made for OoD data synthesizing or collecting.

\textbf{Effects of different watermark injection rounds.} We conduct an ablation study to show the effect of the injection round of the watermark in \cref{tab:earlier_round}. The results verify that injecting in earlier rounds will not affect standard accuracy, WSR, and TAcc. In our paper, we do not start our watermark injection at the very beginning of training since early-stage protection usually means more computational resources, so it is more valuable to focus on high-quality models rather than low-quality models.
\begin{table}[htbp]
    \small
    \centering
    \begin{tabular}{cccccc}
    \toprule
         Inject round&Acc&$\Delta$Acc&WSR&WSR\_Gap&TAcc  \\
         \hline
        5&0.8838&0.0251&0.9951&0.9948&1.0000\\
        10&0.8811&0.0278&0.9946&0.9938&1.0000\\
        20&0.8855&0.0234&0.9909&0.9895&1.0000\\
        
         \bottomrule
    \end{tabular}
    \caption{Ablation study: results for watermark injection in different rounds on digits.}
    \label{tab:earlier_round}
    \vspace{-0.25in}
\end{table}

\begin{figure}[htbp!]
    \centering
    \begin{subfigure}  {0.32\textwidth}\centerline{\includegraphics[width=4.6cm]{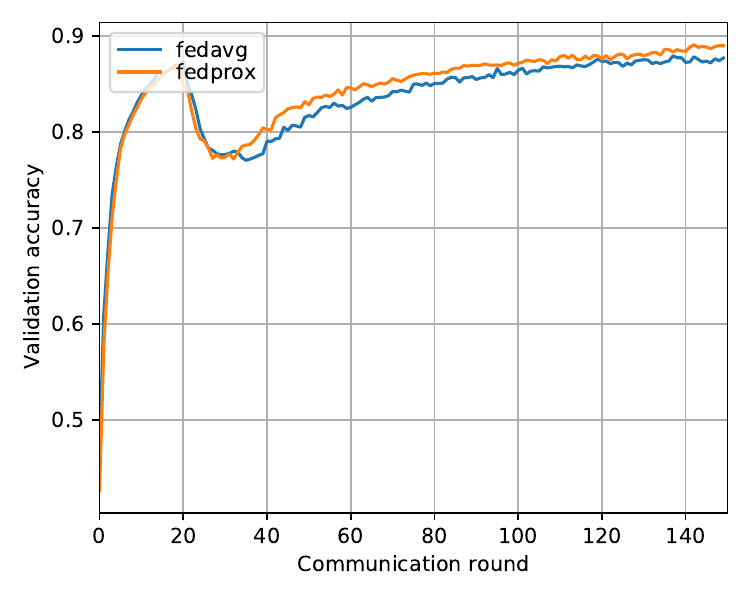}}
    \vspace{-0.15in}
        \subcaption{Validation accuracy.}
    \end{subfigure}
     \begin{subfigure}  {0.32\textwidth}\centerline{\includegraphics[width=4.6cm]{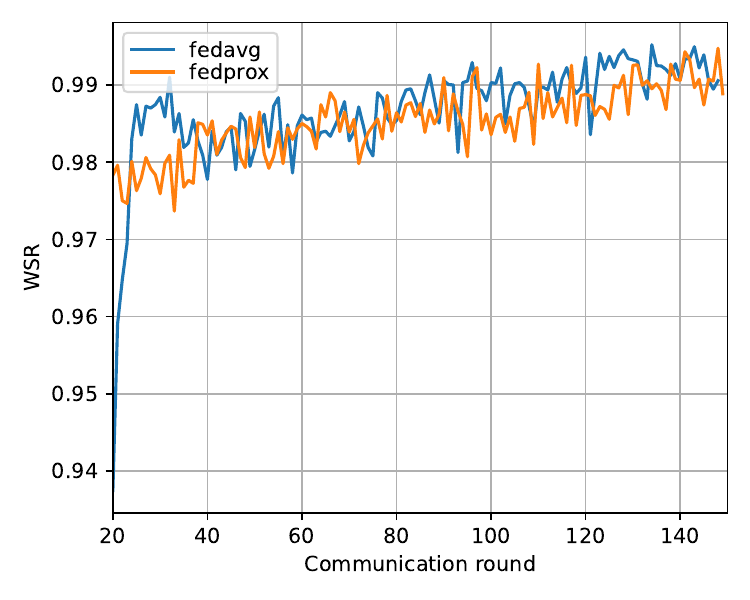}}
     \vspace{-0.15in}
        \subcaption{WSR.}
    \end{subfigure}
     \begin{subfigure}  {0.32\textwidth}\centerline{\includegraphics[width=4.6cm]{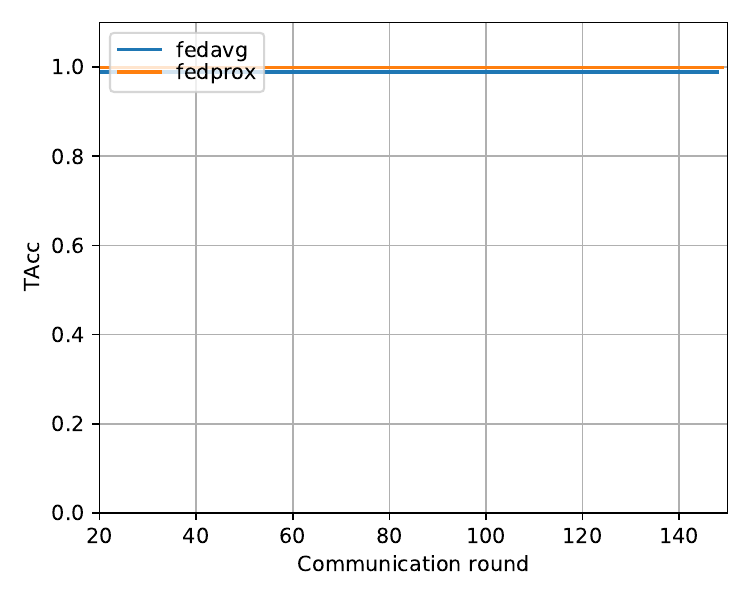}}
     \vspace{-0.15in}
        \subcaption{TAcc.}
    \end{subfigure}
    \caption{Validation accuracy, WSR, and TAcc for fedavg and fedprox on digits.
    }
    \vspace{-0.15in}
    \label{fig:diff_fl}
\end{figure}

\textbf{Effects of different FL algorithms}
In \cref{fig:diff_fl}, we show the standard accuracy, WSR and TAcc for our proposed DUW in two different FL settings: fedavg and fedprox~\citep{li2020federated}. According to the results, fedprox can achieve comparable WSR as fedavg and higher standard accuracy. TAcc for both FL algorithms remains to be $100\%$. Our proposed method is not sensitive to the FL framework, in which it is implanted.
\subsection{Extended robustness study}
\begin{figure}[htbp!]
    \centering
    \includegraphics[width=8cm]{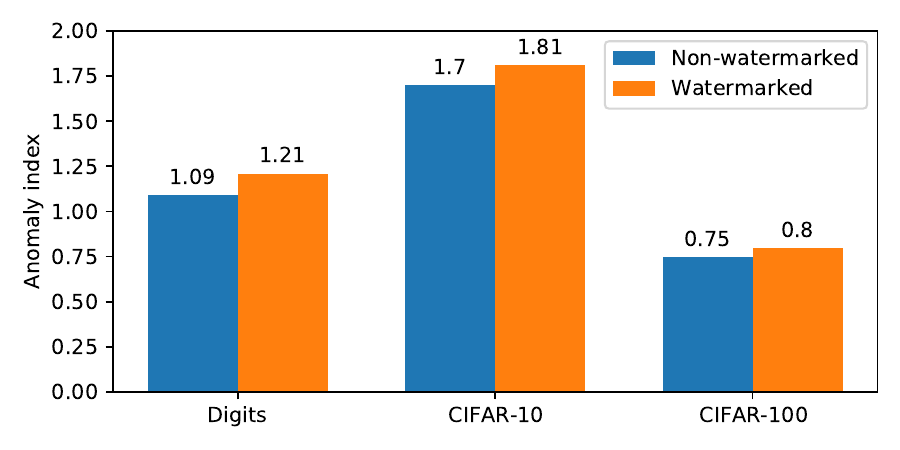}
    \caption{Anomaly index of watermarked model and non-watermarked model. If the anomaly index exceeds $2$, the model will be detected as backdoor-based watermarked.}
    \label{fig:nc_index}
\end{figure}

\textbf{Robustness against detection attack.}
We take Neural Cleanse~\citep{wang2019neural} as an example of the detection attack, which synthesizes the possible trigger to convert all benign images to all possible target classes in the classification task space.  Then anomaly detection is conducted to detect if any trigger
candidate is significantly smaller than other candidates. 
We follow the original setting in~\cite{wang2019neural}, if the anomaly index is larger than $2$, the model is watermarked. The smaller the value of the anomaly index, the harder the watermark to be washed out by Neural Cleanse.  Local samples are used as benign images during detection. 
We compare the anomaly index for
the non-watermarked model and watermarked model in \cref{fig:nc_index}. We observe that for all datasets, the anomaly index for the watermarked model is close to that of the non-watermarked model, and both of them are smaller than the threshold $2$. The observation implies that our watermarked model cannot be detected using neural cleanse. One possible reason is that Neural Cleanse relies on the assumption that the backdoor-based watermark shares the same task space with the original classification task, but due to the effectiveness of our decoder, 
our target label space of the watermark is different from the original classification task space. Therefore, by searching all possible target classes in the original task space, Neural Cleanse will not find the real target label of the trigger set introduced by our proposed DUW.

\begin{figure}[htbp!]
    \centering
    \begin{subfigure}{0.15\textwidth}\centerline{\includegraphics[width=2cm]{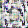}}
        \subcaption{Digits: \\non-wm.}
        \end{subfigure}  
        \begin{subfigure}{0.15\textwidth}
            \centerline{\includegraphics[width=2cm]{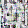}}
        \subcaption{Digits: \\wm.}
        
        \end{subfigure}  
            \begin{subfigure}{0.15\textwidth}
            \centerline{\includegraphics[width=2cm]{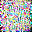}}
        \subcaption{CIFAR-10:\\ non-wm.}
        \end{subfigure}  
          \begin{subfigure} {0.15\textwidth}
            \centerline{\includegraphics[width=2cm]{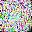}}
        \subcaption{CIFAR-10:\\ wm. }
        \end{subfigure}  
              \begin{subfigure} {0.16\textwidth}
            \centerline{\includegraphics[width=2cm]{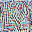}}
        \subcaption{CIFAR-100:\\ non-wm.}
        \end{subfigure}  
                     \begin{subfigure} {0.16\textwidth}
            \centerline{\includegraphics[width=2cm]{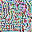}}
        \subcaption{CIFAR-100:\\ wm.}
        \end{subfigure}  
    \caption{Reversed trigger patterns generated by Neural Cleanse for non-watermarked (non-wm) and watermarked (wm) models on three benchmarks.
    }
    \label{fig:nc_trigger}
\end{figure}

We further show the reversed trigger pattern generated by Neural Cleanse for non-watermarked (non-wm) and watermarked (wm) models in \cref{fig:nc_trigger}. The reversed trigger of our watermarked model shares a similar pattern as non-watermarked ones for all three benchmarks, and it does not look similar to our real trigger patterns (real ones can be referred to \cref{fig:vis_ood} residual). The trigger patterns for our trigger sets are sample-specific. Thus, it is hard to reverse engineer triggers when Neural Cleanse assumes a general trigger pattern for the entire trigger set. In summary, our proposed DUW is secured against this trigger-detection algorithm. 

%% file: iclr2024_conference.bbl
\begin{thebibliography}{43}
\providecommand{\natexlab}[1]{#1}
\providecommand{\url}[1]{\texttt{#1}}
\expandafter\ifx\csname urlstyle\endcsname\relax
  \providecommand{\doi}[1]{doi: #1}\else
  \providecommand{\doi}{doi: \begingroup \urlstyle{rm}\Url}\fi

\bibitem[Adi et~al.(2018)Adi, Baum, Cisse, Pinkas, and Keshet]{adi2018turning}
Yossi Adi, Carsten Baum, Moustapha Cisse, Benny Pinkas, and Joseph Keshet.
\newblock Turning your weakness into a strength: Watermarking deep neural networks by backdooring.
\newblock In \emph{27th $\{$USENIX$\}$ Security Symposium ($\{$USENIX$\}$ Security 18)}, pp.\  1615--1631, 2018.

\bibitem[Bagdasaryan et~al.(2020)Bagdasaryan, Veit, Hua, Estrin, and Shmatikov]{bagdasaryan2020backdoor}
Eugene Bagdasaryan, Andreas Veit, Yiqing Hua, Deborah Estrin, and Vitaly Shmatikov.
\newblock How to backdoor federated learning.
\newblock In \emph{International conference on artificial intelligence and statistics}, pp.\  2938--2948. PMLR, 2020.

\bibitem[Caldas et~al.(2018)Caldas, Duddu, Wu, Li, Kone{\v{c}}n{\`y}, McMahan, Smith, and Talwalkar]{caldas2018leaf}
Sebastian Caldas, Sai Meher~Karthik Duddu, Peter Wu, Tian Li, Jakub Kone{\v{c}}n{\`y}, H~Brendan McMahan, Virginia Smith, and Ameet Talwalkar.
\newblock Leaf: A benchmark for federated settings.
\newblock \emph{arXiv preprint arXiv:1812.01097}, 2018.

\bibitem[Chen et~al.(2021)Chen, Chen, Zhang, and Wang]{chen2021you}
Xuxi Chen, Tianlong Chen, Zhenyu Zhang, and Zhangyang Wang.
\newblock You are caught stealing my winning lottery ticket! making a lottery ticket claim its ownership.
\newblock \emph{Advances in Neural Information Processing Systems}, 34:\penalty0 1780--1791, 2021.

\bibitem[Chrabaszcz et~al.(2017)Chrabaszcz, Loshchilov, and Hutter]{chrabaszcz2017downsampled}
Patryk Chrabaszcz, Ilya Loshchilov, and Frank Hutter.
\newblock A downsampled variant of imagenet as an alternative to the cifar datasets.
\newblock \emph{arXiv preprint arXiv:1707.08819}, 2017.

\bibitem[Coates et~al.(2011)Coates, Ng, and Lee]{coates2011analysis}
Adam Coates, Andrew Ng, and Honglak Lee.
\newblock An analysis of single-layer networks in unsupervised feature learning.
\newblock In \emph{Proceedings of the fourteenth international conference on artificial intelligence and statistics}, pp.\  215--223. JMLR Workshop and Conference Proceedings, 2011.

\bibitem[Darvish~Rouhani et~al.(2019)Darvish~Rouhani, Chen, and Koushanfar]{darvish2019deepsigns}
Bita Darvish~Rouhani, Huili Chen, and Farinaz Koushanfar.
\newblock Deepsigns: An end-to-end watermarking framework for ownership protection of deep neural networks.
\newblock In \emph{Proceedings of the Twenty-Fourth International Conference on Architectural Support for Programming Languages and Operating Systems}, pp.\  485--497, 2019.

\bibitem[Fan et~al.(2019)Fan, Ng, and Chan]{fan2019rethinking}
Lixin Fan, Kam~Woh Ng, and Chee~Seng Chan.
\newblock Rethinking deep neural network ownership verification: Embedding passports to defeat ambiguity attacks.
\newblock \emph{Advances in neural information processing systems}, 32, 2019.

\bibitem[Ganin \& Lempitsky(2015)Ganin and Lempitsky]{ganin2015unsupervised}
Yaroslav Ganin and Victor Lempitsky.
\newblock Unsupervised domain adaptation by backpropagation.
\newblock In \emph{International conference on machine learning}, pp.\  1180--1189. PMLR, 2015.

\bibitem[Garg et~al.(2020)Garg, Kumar, Goel, and Liang]{garg2020can}
Siddhant Garg, Adarsh Kumar, Vibhor Goel, and Yingyu Liang.
\newblock Can adversarial weight perturbations inject neural backdoors.
\newblock In \emph{Proceedings of the 29th ACM International Conference on Information \& Knowledge Management}, pp.\  2029--2032, 2020.

\bibitem[Goldblum et~al.(2022)Goldblum, Tsipras, Xie, Chen, Schwarzschild, Song, M{\k{a}}dry, Li, and Goldstein]{goldblum2022dataset}
Micah Goldblum, Dimitris Tsipras, Chulin Xie, Xinyun Chen, Avi Schwarzschild, Dawn Song, Aleksander M{\k{a}}dry, Bo~Li, and Tom Goldstein.
\newblock Dataset security for machine learning: Data poisoning, backdoor attacks, and defenses.
\newblock \emph{IEEE Transactions on Pattern Analysis and Machine Intelligence}, 45\penalty0 (2):\penalty0 1563--1580, 2022.

\bibitem[Gu et~al.(2019)Gu, Liu, Dolan-Gavitt, and Garg]{gu2019badnets}
Tianyu Gu, Kang Liu, Brendan Dolan-Gavitt, and Siddharth Garg.
\newblock Badnets: Evaluating backdooring attacks on deep neural networks.
\newblock \emph{IEEE Access}, 7:\penalty0 47230--47244, 2019.

\bibitem[He et~al.(2016)He, Zhang, Ren, and Sun]{he2016deep}
Kaiming He, Xiangyu Zhang, Shaoqing Ren, and Jian Sun.
\newblock Deep residual learning for image recognition.
\newblock In \emph{Proceedings of the IEEE conference on computer vision and pattern recognition}, pp.\  770--778, 2016.

\bibitem[Hong et~al.(2022)Hong, Wang, Wang, and Zhou]{hong2022efficient}
Junyuan Hong, Haotao Wang, Zhangyang Wang, and Jiayu Zhou.
\newblock Efficient split-mix federated learning for on-demand and in-situ customization.
\newblock \emph{arXiv preprint arXiv:2203.09747}, 2022.

\bibitem[Hull(1994)]{hull1994database}
Jonathan~J. Hull.
\newblock A database for handwritten text recognition research.
\newblock \emph{IEEE Transactions on pattern analysis and machine intelligence}, 16\penalty0 (5):\penalty0 550--554, 1994.

\bibitem[Kone{\v{c}}n{\`y} et~al.(2015)Kone{\v{c}}n{\`y}, McMahan, and Ramage]{konevcny2015federated}
Jakub Kone{\v{c}}n{\`y}, Brendan McMahan, and Daniel Ramage.
\newblock Federated optimization: Distributed optimization beyond the datacenter.
\newblock \emph{arXiv preprint arXiv:1511.03575}, 2015.

\bibitem[Kone{\v{c}}n{\`y} et~al.(2016)Kone{\v{c}}n{\`y}, McMahan, Yu, Richt{\'a}rik, Suresh, and Bacon]{konevcny2016federated}
Jakub Kone{\v{c}}n{\`y}, H~Brendan McMahan, Felix~X Yu, Peter Richt{\'a}rik, Ananda~Theertha Suresh, and Dave Bacon.
\newblock Federated learning: Strategies for improving communication efficiency.
\newblock \emph{arXiv preprint arXiv:1610.05492}, 2016.

\bibitem[Krizhevsky et~al.(2009)Krizhevsky, Hinton, et~al.]{krizhevsky2009learning}
Alex Krizhevsky, Geoffrey Hinton, et~al.
\newblock Learning multiple layers of features from tiny images.
\newblock 2009.

\bibitem[Kuribayashi et~al.(2021)Kuribayashi, Tanaka, Suzuki, Yasui, and Funabiki]{kuribayashi2021white}
Minoru Kuribayashi, Takuro Tanaka, Shunta Suzuki, Tatsuya Yasui, and Nobuo Funabiki.
\newblock White-box watermarking scheme for fully-connected layers in fine-tuning model.
\newblock In \emph{Proceedings of the 2021 ACM Workshop on Information Hiding and Multimedia Security}, pp.\  165--170, 2021.

\bibitem[Le~Merrer et~al.(2020)Le~Merrer, Perez, and Tr{\'e}dan]{le2020adversarial}
Erwan Le~Merrer, Patrick Perez, and Gilles Tr{\'e}dan.
\newblock Adversarial frontier stitching for remote neural network watermarking.
\newblock \emph{Neural Computing and Applications}, 32:\penalty0 9233--9244, 2020.

\bibitem[LeCun et~al.(1998)LeCun, Bottou, Bengio, and Haffner]{lecun1998gradient}
Yann LeCun, L{\'e}on Bottou, Yoshua Bengio, and Patrick Haffner.
\newblock Gradient-based learning applied to document recognition.
\newblock \emph{Proceedings of the IEEE}, 86\penalty0 (11):\penalty0 2278--2324, 1998.

\bibitem[Li et~al.(2021{\natexlab{a}})Li, Wang, and Liew]{li2021towards}
Fang-Qi Li, Shi-Lin Wang, and Alan Wee-Chung Liew.
\newblock Towards practical watermark for deep neural networks in federated learning.
\newblock \emph{arXiv preprint arXiv:2105.03167}, 2021{\natexlab{a}}.

\bibitem[Li et~al.(2022)Li, Yang, Wang, and Liew]{li2022leveraging}
Fangqi Li, Lei Yang, Shilin Wang, and Alan Wee-Chung Liew.
\newblock Leveraging multi-task learning for umambiguous and flexible deep neural network watermarking.
\newblock In \emph{SafeAI@ AAAI}, 2022.

\bibitem[Li et~al.(2020)Li, Sahu, Zaheer, Sanjabi, Talwalkar, and Smith]{li2020federated}
Tian Li, Anit~Kumar Sahu, Manzil Zaheer, Maziar Sanjabi, Ameet Talwalkar, and Virginia Smith.
\newblock Federated optimization in heterogeneous networks.
\newblock \emph{Proceedings of Machine learning and systems}, 2:\penalty0 429--450, 2020.

\bibitem[Li et~al.(2021{\natexlab{b}})Li, Jiang, Zhang, Kamp, and Dou]{li2021fedbn}
Xiaoxiao Li, Meirui Jiang, Xiaofei Zhang, Michael Kamp, and Qi~Dou.
\newblock Fedbn: Federated learning on non-iid features via local batch normalization.
\newblock \emph{arXiv preprint arXiv:2102.07623}, 2021{\natexlab{b}}.

\bibitem[Li et~al.(2021{\natexlab{c}})Li, Li, Wu, Li, He, and Lyu]{li2021invisible}
Yuezun Li, Yiming Li, Baoyuan Wu, Longkang Li, Ran He, and Siwei Lyu.
\newblock Invisible backdoor attack with sample-specific triggers.
\newblock In \emph{Proceedings of the IEEE/CVF International Conference on Computer Vision}, pp.\  16463--16472, 2021{\natexlab{c}}.

\bibitem[Liang \& Wang(2023)Liang and Wang]{liang2023fedcip}
Junchuan Liang and Rong Wang.
\newblock Fedcip: Federated client intellectual property protection with traitor tracking.
\newblock \emph{arXiv preprint arXiv:2306.01356}, 2023.

\bibitem[Liu et~al.(2018)Liu, Sun, Zhou, Huang, and Darrell]{liu2018rethinking}
Zhuang Liu, Mingjie Sun, Tinghui Zhou, Gao Huang, and Trevor Darrell.
\newblock Rethinking the value of network pruning.
\newblock \emph{arXiv preprint arXiv:1810.05270}, 2018.

\bibitem[Marfoq et~al.(2020)Marfoq, Xu, Neglia, and Vidal]{marfoq2020throughput}
Othmane Marfoq, Chuan Xu, Giovanni Neglia, and Richard Vidal.
\newblock Throughput-optimal topology design for cross-silo federated learning.
\newblock \emph{Advances in Neural Information Processing Systems}, 33:\penalty0 19478--19487, 2020.

\bibitem[McMahan et~al.(2017)McMahan, Moore, Ramage, Hampson, and y~Arcas]{mcmahan2017communication}
Brendan McMahan, Eider Moore, Daniel Ramage, Seth Hampson, and Blaise~Aguera y~Arcas.
\newblock Communication-efficient learning of deep networks from decentralized data.
\newblock In \emph{Artificial intelligence and statistics}, pp.\  1273--1282. PMLR, 2017.

\bibitem[Mehta et~al.(2022)Mehta, Mondol, Farahmandi, and Tehranipoor]{mehta2022aime}
Dhwani Mehta, Nurun Mondol, Farimah Farahmandi, and Mark Tehranipoor.
\newblock Aime: watermarking ai models by leveraging errors.
\newblock In \emph{2022 Design, Automation \& Test in Europe Conference \& Exhibition (DATE)}, pp.\  304--309. IEEE, 2022.

\bibitem[Netzer et~al.(2011)Netzer, Wang, Coates, Bissacco, Wu, and Ng]{netzer2011reading}
Yuval Netzer, Tao Wang, Adam Coates, Alessandro Bissacco, Bo~Wu, and Andrew~Y Ng.
\newblock Reading digits in natural images with unsupervised feature learning.
\newblock 2011.

\bibitem[Orekondy et~al.(2019)Orekondy, Schiele, and Fritz]{orekondy2019knockoff}
Tribhuvanesh Orekondy, Bernt Schiele, and Mario Fritz.
\newblock Knockoff nets: Stealing functionality of black-box models.
\newblock In \emph{Proceedings of the IEEE/CVF conference on computer vision and pattern recognition}, pp.\  4954--4963, 2019.

\bibitem[Shao et~al.(2022)Shao, Yang, Gu, Lou, Qin, Fan, Yang, and Ren]{shao2022fedtracker}
Shuo Shao, Wenyuan Yang, Hanlin Gu, Jian Lou, Zhan Qin, Lixin Fan, Qiang Yang, and Kui Ren.
\newblock Fedtracker: Furnishing ownership verification and traceability for federated learning model.
\newblock \emph{arXiv preprint arXiv:2211.07160}, 2022.

\bibitem[Stallkamp et~al.(2012)Stallkamp, Schlipsing, Salmen, and Igel]{stallkamp2012man}
Johannes Stallkamp, Marc Schlipsing, Jan Salmen, and Christian Igel.
\newblock Man vs. computer: Benchmarking machine learning algorithms for traffic sign recognition.
\newblock \emph{Neural networks}, 32:\penalty0 323--332, 2012.

\bibitem[T~Dinh et~al.(2020)T~Dinh, Tran, and Nguyen]{t2020personalized}
Canh T~Dinh, Nguyen Tran, and Josh Nguyen.
\newblock Personalized federated learning with moreau envelopes.
\newblock \emph{Advances in Neural Information Processing Systems}, 33:\penalty0 21394--21405, 2020.

\bibitem[Tang et~al.(2020)Tang, Du, Liu, Yang, and Hu]{tang2020embarrassingly}
Ruixiang Tang, Mengnan Du, Ninghao Liu, Fan Yang, and Xia Hu.
\newblock An embarrassingly simple approach for trojan attack in deep neural networks.
\newblock In \emph{Proceedings of the 26th ACM SIGKDD International Conference on Knowledge Discovery \& Data Mining}, pp.\  218--228, 2020.

\bibitem[Tekgul et~al.(2021)Tekgul, Xia, Marchal, and Asokan]{tekgul2021waffle}
Buse~GA Tekgul, Yuxi Xia, Samuel Marchal, and N~Asokan.
\newblock Waffle: Watermarking in federated learning.
\newblock In \emph{2021 40th International Symposium on Reliable Distributed Systems (SRDS)}, pp.\  310--320. IEEE, 2021.

\bibitem[Uchida et~al.(2017)Uchida, Nagai, Sakazawa, and Satoh]{uchida2017embedding}
Yusuke Uchida, Yuki Nagai, Shigeyuki Sakazawa, and Shin'ichi Satoh.
\newblock Embedding watermarks into deep neural networks.
\newblock In \emph{Proceedings of the 2017 ACM on international conference on multimedia retrieval}, pp.\  269--277, 2017.

\bibitem[Vincent(2023)]{meta2023}
James Vincent.
\newblock Meta’s powerful ai language model has leaked online — what happens now?
\newblock \url{https://www.theverge.com/2023/3/8/23629362/meta-ai-language-model-llama-leak-online-misuse}, 2023.
\newblock Accessed: 2023-03-08.

\bibitem[Wang et~al.(2019)Wang, Yao, Shan, Li, Viswanath, Zheng, and Zhao]{wang2019neural}
Bolun Wang, Yuanshun Yao, Shawn Shan, Huiying Li, Bimal Viswanath, Haitao Zheng, and Ben~Y Zhao.
\newblock Neural cleanse: Identifying and mitigating backdoor attacks in neural networks.
\newblock In \emph{2019 IEEE Symposium on Security and Privacy (SP)}, pp.\  707--723. IEEE, 2019.

\bibitem[Zhang et~al.(2018)Zhang, Gu, Jang, Wu, Stoecklin, Huang, and Molloy]{zhang2018protecting}
Jialong Zhang, Zhongshu Gu, Jiyong Jang, Hui Wu, Marc~Ph Stoecklin, Heqing Huang, and Ian Molloy.
\newblock Protecting intellectual property of deep neural networks with watermarking.
\newblock In \emph{Proceedings of the 2018 on Asia Conference on Computer and Communications Security}, pp.\  159--172, 2018.

\bibitem[Zhu et~al.(2021)Zhu, Hong, and Zhou]{zhu2021data}
Zhuangdi Zhu, Junyuan Hong, and Jiayu Zhou.
\newblock Data-free knowledge distillation for heterogeneous federated learning.
\newblock In \emph{International conference on machine learning}, pp.\  12878--12889. PMLR, 2021.

\end{thebibliography}
